\documentclass[reprint, NumberedRefs,10pt]{JASA}

\usepackage[shortlabels]{enumitem}

\usepackage{algpseudocode}

\usepackage{tikz}
\usetikzlibrary{calc}
\usetikzlibrary{arrows}
\usetikzlibrary{spy}
\usepackage{relsize}
\usepackage{wrapfig}
\usepackage{eurosym}
\usepackage[siunitx]{circuitikz}
\usepackage{amssymb}
\usepackage{esint}
\usepackage{mathrsfs}
\usepackage{amsthm}
\usepackage{amscd}
\usepackage{listings}
\usepackage{booktabs}
\usepackage{multirow}
\usepackage{textcomp}

\newcommand{\vect}[1]{\boldsymbol{#1}}

\newcommand{\matr}[1]{\mathbf{#1}}
\newcommand{\abs}[1]{\left \lvert #1 \right\rvert }

\renewcommand{\vr}[0]{\mathbf{x}}
\newcommand{\vrp}[0]{\mathbf{y}}
\newcommand{\vrq}[0]{\mathbf{z}}

\newcommand{\matV}[0]{\matr{Y}}
\newcommand{\matJ}[0]{\matr{X}}
\newcommand{\matZ}[0]{\matr{A}}

  \newcommand{\unit}[1]{\vect{\hat{#1}}}

\newcommand{\junk}[1] {}

\def\XXint#1#2#3{{\setbox0=\hbox{$#1{#2#3}{\int}$}
\vcenter{\hbox{$#2#3$}}\kern-.5\wd0}}

\newcommand*\widebar[1]{%
  \hbox{%
    \vbox{%
      \hrule height 0.5pt 
      \kern0.3ex
      \hbox{%
        \kern-0.05em
        \ensuremath{#1}%
        \kern-0.05em
      }%
    }%
  }%
}

\usepackage[siunitx]{circuitikz}
\usepackage{etoolbox}
\makeatletter

\pgfcircdeclarebipole{}
    {\ctikzvalof{bipoles/tline/height}}{tline}
    {\ctikzvalof{bipoles/tline/height}}
    {\ctikzvalof{bipoles/tline/width}}
{   
    \pgfpointdiff{\pgfpointanchor{\ctikzvalof{bipole/name}start}{center}}
                 {\pgfpointanchor{\ctikzvalof{bipole/name}end}{center}}
    \pgfmathparse{veclen(\the\pgf@x,\the\pgf@y)}
    \pgftransformxshift{\pgfmathresult/2-30pt}
    \pgf@circ@res@left=\dimexpr-\pgfmathresult pt+40pt\relax
    \pgf@circ@res@step=.2\pgf@circ@res@right 
    \pgfsetlinewidth{\pgfkeysvalueof{/tikz/circuitikz/bipoles/thickness}\pgfstartlinewidth}
    \pgfpathellipse{\pgfpoint{\pgf@circ@res@right-\pgf@circ@res@step}{0}}
                   {\pgfpoint{\pgf@circ@res@step}{0}}
                   {\pgfpoint{0}{-\pgf@circ@res@up}}
    \pgfpathmoveto{\pgfpoint{\pgf@circ@res@right-\pgf@circ@res@step}{\pgf@circ@res@up}}
    \pgfpathlineto{\pgfpoint{\pgf@circ@res@left+\pgf@circ@res@step}{\pgf@circ@res@up}}
    \pgfpatharc{-90}{90}{-\pgf@circ@res@step and -\pgf@circ@res@up}
    \pgfpathlineto{\pgfpoint{\pgf@circ@res@right-\pgf@circ@res@step}{\pgf@circ@res@down}}
    \pgfsetfillcolor{white}
    \pgfusepath{draw,fill}
    \pgfpathmoveto{\pgfpoint{\pgf@circ@res@right-2*\pgf@circ@res@step}{0pt}}
    \pgfpathlineto{\pgfpoint{\pgf@circ@res@right}{0pt}}
    \pgfusepath{draw}
}

\allowdisplaybreaks

\usepackage{pgfplots}
\usepackage{soul}

\newcommand{\inc}[0]{\text{inc}}
\newcommand{\sca}[0]{\text{sca}}

\begin{document}

\title[Wigner-Smith Time Delay Matrix for Acoustic Scattering: Computational Aspects]{Wigner-Smith Time Delay Matrix for Acoustic Scattering: Computational Aspects}
\author{Utkarsh R. Patel}
\author{Yiqian Mao}
\author{Jack Hamel}
\author{Eric Michielssen}
\affiliation{Department of Electrical Engineering and Computer Science,  University of Michigan, Ann Arbor, Michigan 48109-2122, United States.}

\preprint{Author, JASA}		

\date{\today} 

\begin{abstract}
The Wigner-Smith (WS) time delay matrix relates an acoustic system's scattering matrix to its wavenumber derivative. 
The entries of the WS time delay matrix can be expressed in terms of energy density-like volume integrals, which cannot be efficiently evaluated in a boundary element method framework. 
This paper presents two schemes for efficiently populating the WS time delay matrix. The direct formulation casts the energy density-like volume integrals into integrals of the incident field and the field and/or its normal derivative over the scatterer surface. The indirect formulation computes the system's scattering matrix and its wavenumber derivative, again via surface integration, and then invokes the WS relationship to compute the WS time delay matrix. Both the direct and the indirect formulations yield equivalent results and can be easily integrated into standard boundary element codes.
\end{abstract}


\maketitle

\section{\label{sec:1} Introduction}

Wigner-Smith (WS) techniques were proposed in 1960 by Felix Smith to characterize time delays experienced by particles during quantum mechanical interaction \cite{Smith1960}.
Ever since, they have been applied to a wide range
of problems in quantum mechanics\cite{Buttiker_1982, Wardlaw_1988,Dittes_2000}, photonics\cite{Gallmann_2017,Hockett_2016}, and electromagnetics~\cite{Winful_2003,TAP1_WS,TAP2_WS}. 
An excellent review of WS methods and their applications can
be found in~\cite{Texier_2013}.

This paper is the second by the same authors on the applicability of WS methods to acoustic scattering problems governed by the Helmholtz equation. 
The first paper\cite{WS_Acoustic_Paper1} introduced the  Wigner-Smith (WS) time delay matrix
\begin{align}
    \matr{Q}= j\matr{S}^\dag \frac{d\matr{S}}{dk}\,,
    \label{eq:WS_relation}
\end{align}
where $\matr{S}$ is a system's scattering matrix and $k$ represents the wavenumber. The paper showed that, irrespective of the scatterer's size, shape, and boundary condition (sound-soft or sound-hard), $\matr{Q}$'s entries can be expressed as volume integrals of energy-like quantities.
The paper also discussed a few interesting properties of $\matr{Q}$, 
including the fact that its eigenvectors describe so-called WS modes that have well-defined energies and time delays.

This paper discusses computational aspects of WS methods. Computing $\matr{Q}$ by evaluating the volume integral introduced in the first paper is computationally expensive as it requires performing a three-dimensional integral over $\mathbb{R}^3$. 
This paper presents two approaches to efficiently compute $\matr{Q}$ using the boundary element method (BEM) via surface integration:
\begin{itemize}
    \item In the direct method, the volume integrals are cast into surface integrals whose integrand is a function of the incident fields and the velocity potential.
The resulting expressions are straightforward to evaluate, exhibiting the same computational complexity as those to assemble the BEM matrix, and are trivially integrated into BEM codes.  
 \item In the indirect method, system's scattering matrix and its derivative w.r.t. $k$ are computed first, again via surface integration. 
 $\matr{Q}$ is then computed via its definition Eqn.~\eqref{eq:WS_relation}.
\end{itemize}
It is proven that both the direct and the indirect methods are equivalent.

The paper is organized as follows. Sec. II describes the scattering problem under consideration and a BEM framework for solving it. Sec. III and Sec. IV present direct and indirect methods for computing the WS time delay matrix, respectively. The material in these sections is limited to key equations needed for BEM practitioners to incorporate WS methods in their codes; most technical derivations are relegated to the Appendix and Supplementary Material. Sec. V presents two numerical examples that elucidate the use of WS modes when classifying 3D scattering phenomena. Finally, Sec. VI states the paper's conclusions.

\noindent
\underline{Notation}:
\begin{itemize}[leftmargin=*]
\item $\,^\dag$, $\,^T$, $\,^*$, and $'$ represent adjoint, transpose, complex conjugate, and derivative w.r.t. $k$ (i.e. $d/dk$) operations, respectively. 
\item This paper assumes a $e^{j\omega t}$ dependence with $\omega = v k$ and $v$ is the wave speed.
\item Fields are often expanded in terms of spherical waves ${\cal B}_{lm}(\vr)$. ${\cal B}_{lm}(\vr) = {\cal I}_{lm}(\vr), {\cal O}_{lm}(\vr), {\cal W}_{lm}(\vr)$ when modeling \emph{incoming}, \emph{outgoing}, and \emph{standing} waves.
While \emph{incoming} and \emph{outgoing} waves ${\cal I}_{l m}(\vr)$ and ${\cal O}_{l m}(\vr)$ are singular at the origin, ${\cal W}_{l m}(\vr)$ is regular throughout space.
The large argument approximations (i.e. ``far-fields'') of ${\cal B}_{l m}(\vr)$ are denoted by ${\cal B}_{l m, \infty}(\vr)$.
Expressions and properties of ${\cal B}_{lm}(\vr)$ and ${\cal B}_{lm,\infty}(\vr)$ are provided in Appendix~\ref{Appdix:modes}.
\end{itemize}

\section{Computational Framework}

\begin{figure}
\centering
\includegraphics{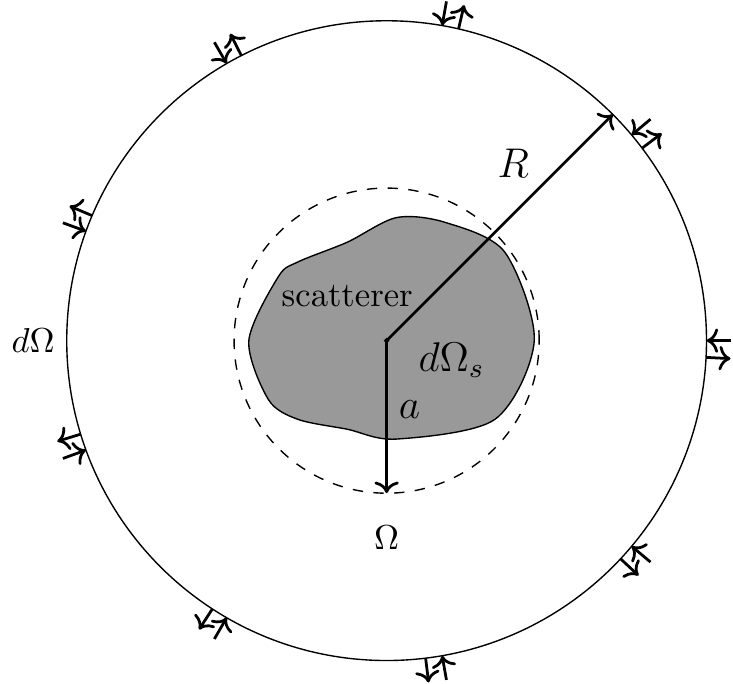}
\caption{Scattering system excited through free-space port defined on sphere of radius $R$.}
\label{fig:sample_scatterer}
\end{figure}

This section presents the acoustic scattering problem under consideration.
It defines incoming and outgoing fields that are used to construct scattering matrices of sound-soft and sound-hard scatterers. 
It also presents integral equations to compute the scattering matrix by decomposing fields into incident and scattered components, along with a procedure to solve these equations via the BEM.

\subsection{Setup and Incoming Fields}

Consider a closed scatterer $\Omega_s$ with surface $d\Omega_s$ that resides in a lossless homogeneous medium with (real) wavenumber $k$ and is circumscribed by a sphere of radius $a$ centered at the origin (Fig.~\ref{fig:sample_scatterer}).
Let $\Omega$ and $d\Omega$ denote the volume and surface of a concentric sphere of radius $R \gg a$ and $kR \gg 1$, respectively. 
The scatterer is excited by $M$ incoming waves. 
For $\vr \in d\Omega$, the incoming field $\phi_{p,\infty}^i(\vr)$ is
\begin{subequations}
\begin{align}
    \phi_{p,\infty}^i(\vr) &=
    {\cal I}_{p,\infty}(\vr)  \label{eq:incoming1} \\ 
    &= \frac{e^{jkr}}{r} {\cal X}_p(\theta,\phi)\,.
    \label{eq:incoming2}
\end{align}
\end{subequations}
Here, ${\cal I}_{p,\infty}(\vr)$ is an incoming spherical wave in the large argument regime of $r \rightarrow \infty$ as defined in the Appendix.
${\cal X}_p(\theta,\phi)$ is a spherical harmonic with $p$ mapping to a tuple $(l,m)$ where $l=0,\hdots, l_{max}$, $m = -l,\hdots, l$, and $l_{max} = k a + c(ka)^{1/3}$ where typically $2 < c< 4$ \cite{Wiscombe1980}.
This choice of $l_{max}$ guarantees that for $l > l_{max}$ the field decays appreciably before it reaches the scatterer.
Note that spherical harmonics obey the orthonormality property
\begin{align}
    \int_{0}^{2\pi} \int_{0}^{\pi} {\cal X}_p(\theta,\phi) {\cal X}_{p'}^*(\theta,\phi) \sin \theta d\theta d\phi = \delta_{pp'} \label{eq:X_p_orthonorm}
\end{align}
where $\delta_{pp'}$ is the Kronecker delta function.

\subsection{Outgoing Fields and Scattering Matrices}

The total field at $\vr \in d\Omega$ is the superposition of incoming and outgoing fields.
The outgoing field near $d\Omega$ can be expressed as
\begin{subequations}
\begin{align}
    \phi_{p,\infty}^o(\vr) &= \sum_{t=1}^{M} \matr{S}_{tp} {\cal I}_{t,\infty}^*(\vr) \label{eq:outgoing1} \\
    &= \sum_{t=1}^{M} \matr{S}_{tp}    \frac{e^{-jkr}}{r} {\cal X}_{t}^*(\theta,\phi) \label{eq:outgoing2}
\end{align}
\end{subequations}
where ${\cal I}_{p}^*(\vr) = (-1)^{1+l+m} {\cal O}_{\tilde{p}}(\vr)$ is the outgoing spherical wave, $\tilde{p} = (l,-m)$, and $\matr{S}_{tp}$ is the scattering coefficient representing coupling from mode $p$ to mode $t$.
Representing the outgoing field in terms of ${\cal I}_{p}^*(\vr)$ instead of ${\cal O}_p(\vr)$ ensures that the scattering matrix is both unitary ($\matr{S}^\dag \matr{S} = \matr{I}$) and symmetric $(\matr{S} = \matr{S}^T$).

From Eqns.~\eqref{eq:incoming1}, \eqref{eq:incoming2}, \eqref{eq:outgoing1}, and \eqref{eq:outgoing2} the total field for $\vr \in d\Omega$ reads
\begin{subequations}
\begin{align}
    &\phi_{p,\infty}(\vr) = \phi_{p,\infty}^i(\vr) + \phi_{p,\infty}^o(\vr) \\
    &\quad = {\cal I}_{p,\infty}(\vr) + \sum_{t=1}^M \matr{S}_{tp} {\cal I}_{t,\infty}^*(\vr) \label{eq:phitot_Dirichlet} \\
    &\quad = \frac{e^{jkr}}{r} {\cal X}_p(\theta,\phi) + \sum_{t=1}^M \matr{S}_{tp} \frac{e^{-jkr}}{r} {\cal X}_t^*(\theta,\phi)\,.
\end{align}
\end{subequations}
Inside $\Omega$, the total field due to $\phi_{p,\infty}^i(\vr)$ is denoted by $\phi_p(\vr)$.

\subsection{Incident-Scattered Field Decomposition}

While the above decomposition of the total field into incoming and outgoing components is useful in defining the scattering matrix, it is inconvenient to implement in a BEM framework. 
To facilitate the latter, consider the ``incident field'' $\phi_{p}^{\text{inc}}(\vr)$ that arises when the incoming field $\phi_{p,\infty}^i$ is injected into $\Omega$ in the absence of the scatterer $d\Omega_s$, and decompose the total field $\phi_p(\vr)$ as
\begin{align}
    \phi_p(\vr) &= \phi_{p}^{\text{inc}}(\vr) + \phi_{p}^{\text{sca}}(\vr)
    \label{eq:phip_tot_inc_sca}
\end{align}
where $\phi_{p}^{\text{sca}}(\vr)$ denotes the ``scattered field".
In the absence of a scatterer, the incoming field ``reflects'' upon reaching the origin producing a standing wave that for $\vr \in \Omega$ equals
\begin{subequations}
\begin{align}
    \phi_p^\inc(\vr) &= {\cal W}_p(\vr) \label{eq:phi_inc}\\
    &= {\cal I}_{lm}(\vr) + {\cal O}_{lm}(\vr)  \\
    &= 2 k j^{l+1}  j_p(k r) {\cal X}_{lm}(\theta,\phi) \,.
\end{align}
\end{subequations}
The spherical standing waves ${\cal W}_{p}(\vr)$ are regular at the origin. Expressions for the scattered field $\phi_{p}^\sca(\vr)$ in Eqn.~\eqref{eq:phip_tot_inc_sca} depend on the boundary condition enforced on the scatterer.~\cite{Kirkup2007}\\

\noindent
\underline{Sound-Soft}: For sound-soft scatterers, the scattered field at $\vr \in \Omega$ reads
    \begin{subequations}
    \begin{align}
        \phi_p^{\sca}(\vr) &= {\cal L} \left[\sigma_p\right](\vr) \label{eq:Loperator}\\
        &= \int_{d\Omega_{s}} \sigma_p(\vrp) G(\vr,\vrp) d\vrp
    \end{align}
    \end{subequations}
    where the density $\sigma_p(\vr) = -\frac{\partial}{\partial n'} \phi_p(\vr)$ on $d\Omega_{s}$ and
     $G(\vr,\vrp) = \frac{e^{-jk\abs{\vr-\vrp}}}{4\pi \abs{\vr-\vrp}}$
    is the Green's function of the Helmholtz equation.
    
    \noindent
\underline{Sound-Hard}: For sound-hard scatterers, the scattered field at $\vr \in \Omega$ reads
\begin{subequations}
\begin{align}
&\phi_p^\sca(\vr) = {\cal K} \left[\sigma_p\right](\vr) \label{eq:Koperator}\\
&= \frac{1}{2} \sigma_p(\vr) + \mathrm{p.v.}\int_{d\Omega_{s}} \sigma_p(\vrp) \frac{\partial}{\partial n'} G(\vr,\vrp) d\vrp
\label{eq:Koperator2}
\end{align}
\end{subequations}
where $\sigma_p(\vr) = \phi_p(\vr)$ on $d\Omega_s$, $\partial/\partial n'$ denotes the normal derivative, and $\text{p.v.}$ represents principal value. 


\subsection{Integral Equations}

Substituting Eqns.~\eqref{eq:Loperator}, \eqref{eq:Koperator}, and \eqref{eq:phi_inc} into Eqn.~\eqref{eq:phip_tot_inc_sca} yields integral equations for the source density $\sigma_p(\vr)$.\\

\noindent
\underline{Sound-Soft}: For sound-soft scatterers, enforcing $\phi_p(\vr) = 0$ on $d\Omega_s$ produces
the resonant-free combined field integral equation (CFIE) 
    \begin{align}
        (1-\alpha) &{\cal L}\left[\sigma_p\right](\vr) + \alpha \frac{1}{2} \sigma_p + \alpha \, {\cal K}_t\left[\sigma_p\right](\vr) \nonumber \\
        &= -(1-\alpha)\phi_p^\inc(\vr) -\alpha \frac{\partial}{\partial n}\phi_p^\inc(\vr)
        \label{eq:Dirichlet_IE}
    \end{align}
where 
 \begin{align}
        {\cal K}_t\left[\sigma_p\right](\vr) &=
        \text{p.v.} \left[ \frac{\partial}{\partial n}{\cal L}\left[\sigma_p\right](\vr)\right]
    \end{align}
and $\alpha$ is a constant.\\

\noindent 
\underline{Sound-Hard}: For sound-hard scatterers, enforcing $\partial \phi_p(\vr)/\partial n = 0$ yields
    the CFIE
    \begin{align}
     - \frac{1}{2} &\alpha \sigma_p(\vr) + \alpha {\cal K}\left[\sigma_p\right](\vr) +   (1-\alpha) {\cal M} \left[\sigma_p\right](\vr) \nonumber \\
     &= -\alpha \phi_p^\inc(\vr) - (1-\alpha) \frac{\partial \phi_p^\inc(\vr)}{\partial n}
     \label{eq:cfie_neumann}
    \end{align}
    where
    \begin{align}
        {\cal M}\left[\sigma_p\right](\vr) &= \int_{d\Omega_s} \sigma_p(\vrp) \frac{\partial}{\partial n} \frac{\partial}{\partial n'} G(\vr,\vrp) d\vrp\,. 
        \label{eq:cfie_neumann_M}
    \end{align}
    Eqn.~\eqref{eq:cfie_neumann} is also known as the Burton-Miller integral equation \cite{Burton1971app}.

\subsection{BEM Implementation}

Integral equations \eqref{eq:Dirichlet_IE} and \eqref{eq:cfie_neumann} can be solved by the BEM. \\

\noindent
\underline{Sound-Soft}: 
Suppose that $\sigma_p(\vr)$ is discretized with real-valued basis functions $f_n(\vr)$
\begin{align}
    \sigma_p(\vr) = \sum_{n=1}^{N} \sigma_{n,p} f_n(\vr)\,. 
    \label{eq:phi_ps_expand_dirichlet}
\end{align}
The derivations in this section do not assume a specific choice for the $f_n(\vr)$; the numerical result section assumes they are pulses defined on a triangular mesh. Substituting \eqref{eq:phi_ps_expand_dirichlet} into \eqref{eq:Dirichlet_IE} yields
\begin{align}
    \widehat{\matZ} \matJ &= \widehat{\matV} 
    \label{eq:CFIE_discretized_Dirichlet}
\end{align}
where $\matJ = \begin{bmatrix} \matJ_1 & \hdots & \matJ_2 \end{bmatrix}$ is matrix of unknowns with $\matJ_{p} = \begin{bmatrix} \sigma_{1,p} & \hdots & \sigma_{N,p} \end{bmatrix}^T$, $\widehat{\matZ} = (1-\alpha) \matZ + \alpha \widetilde{\matZ}$ is the $N \times N$ impedance matrix, and $\widehat{\matV} = \matV + \alpha \widetilde{\matV}$ 
is the $N \times M$ matrix whose $p$-th column corresponds to excitation from the $p$-th mode.
The $(m,n)$-th elements of $\matZ$ and $\widetilde{\matZ}$ are
\begin{subequations}
\begin{align}
    \matZ_{mn} &= \int_{d\Omega_s} f_m(\vr) {\cal L}\left[f_n \right](\vr) d\vr  \\
     \widetilde{\matZ}_{mn} &=  \int_{d\Omega_s} f_m(\vr) \left[ \frac{1}{2} f_n(\vr) + {\cal K}_t \left[f_n\right](\vr) \right] d\vr\,. 
    \end{align}
\end{subequations}
The $(n,p)$th entries of $\matV$ and $\widetilde{\matV}$ read
\begin{subequations}
\begin{align}
    \matV_{np} &= - \int_{d\Omega_s} \phi_p^\inc(\vr) f_n(\vr) d\vr \label{eq:Vs_np}\\
    \widetilde{\matV}_{np} &= - \int_{d\Omega_s} \bigg[ \frac{\partial}{\partial n} \phi_p^\inc(\vr) \bigg] f_n(\vr) d\vr\,.
\end{align}
\end{subequations}
Solving the linear system in Eqn.~\eqref{eq:CFIE_discretized_Dirichlet} yields the  values of unknowns $\sigma_{n,p}$ for $n=1,\hdots, N$ and $p=1,\hdots, M$.\\

\noindent
\underline{Sound-Hard}: Assuming that $\sigma_p(\vr)$ is expanded as in Eqn.~\eqref{eq:phi_ps_expand_dirichlet}, Eqn.~\eqref{eq:cfie_neumann} can be discretized resulting in matrix equation Eqn.~\eqref{eq:CFIE_discretized_Dirichlet}, but now with elements
\begin{subequations}
\begin{align}
    \matZ_{mn} &= 
    \int_{d\Omega_s} f_m(\vr) {\cal M}\left[f_n\right](\vr) d\vr \label{eq:Zmn_Neumann}\\
     \widetilde{\matZ}_{mn} &=  \int_{d\Omega_s}  f_m(\vr) \left[-\frac{1}{2}f_n(\vr) + {\cal K} \left[f_n\right](\vr) \right] d\vr 
\end{align}
\end{subequations}
and
\begin{subequations}
\begin{align}
    \matV_{np} &= - \int_{d\Omega_s} \frac{\partial \phi_p^\inc(\vr)}{\partial n} f_n(\vr) d\vr \label{eq:Vh_Neumann}\\
    \widetilde{\matV}_{np} &= - \int_{d\Omega_s}  \phi_p^\inc(\vr)  f_n(\vr) d\vr\,.
\end{align}
\end{subequations}

\section{Direct Computation of $\matr{Q}$ via Integration of Energy-Density Like Quantities}
\label{sec:direct_Q}

As discussed in \cite{WS_Acoustic_Paper1}, the elements of $\matr{Q}$ can be obtained via the computationally expensive evaluation of the following volume integral:
\begin{align}
    \matr{Q}_{qp} &= \frac{1}{2} \int_{\Omega} \left[\phi_p(\vr) \phi_q^*(\vr) - \phi_{p,\infty}(\vr) \phi_{q,\infty}^*(\vr) \right]d\vr \nonumber \\
    &\quad + \frac{1}{2k^2} \int_{\Omega} \big[\nabla \phi_q^*(\vr) \cdot \nabla \phi_p(\vr) \nonumber \\
    &\quad \quad \quad - \nabla \phi_{q,\infty}^*(\vr) \cdot \nabla \phi_{p,\infty}(\vr) \big] d\vr\,. \label{eq:Qvolumen1}
\end{align}
 A more efficient method for evaluating the above integral using surface integral operators acting on $\sigma_p(\vr)$,  $\phi_p^\inc(\vr)$, and $\partial \phi_p^\inc(\vr)/\partial n$ is presented next. A procedure for discretizing these operators in the aforementioned BEM framework is outlined as well.

\subsection{Surface Integral Expressions of the WS Time Delay Matrix}

Substituting the incident-scattered field decomposition of $\phi_p(\vr)$ and $\phi_q(\vr)$ into Eqn.~\eqref{eq:Qvolumen1} yields
\begin{align}
    \matr{Q}_{qp} &= \matr{Q}_{qp}^{\text{inc,inc}} + \matr{Q}_{qp}^{\text{sca,inc}} + \matr{Q}_{qp}^{\text{inc,sca}} + \matr{Q}_{qp}^{\text{sca,sca}}\,
    \label{eq:Qincsca_decompose}
\end{align}
where for $\alpha, \beta \in \{\text{inc},\text{sca} \}$
\begin{align}
 \matr{Q}_{qp}^{\alpha,\beta} &= \frac{1}{2} \int_{\Omega} \left[\phi_p^\beta (\vr) \phi_q^{\alpha *}(\vr) - \phi_{p,\infty}^{\beta}(\vr) \phi_{q,\infty}^{\alpha *}(\vr) \right]d\vr \nonumber \\
 &\quad + \frac{1}{2k^2} \int_{\Omega} \bigg[\nabla \phi_q^{\alpha *}(\vr) \cdot \nabla \phi_p^{\beta}(\vr) \nonumber \\
 &\quad \quad \quad - \nabla \phi_{q,\infty}^{\alpha *}(\vr) \cdot \nabla \phi_{p,\infty}^{\beta} (\vr) \bigg] d\vr\,.
 \label{eq:Qalphabeta_volumeintegral}
\end{align}
A lengthy derivation presented in the Appendix~\ref{app:direct_comp_of_Q} yields surface integral expressions for the four integrals in Eqn.~\eqref{eq:Qincsca_decompose}.\\

\noindent
\underline{Sound-Soft}:
 For sound-soft scatterers, the four terms in Eqn.~\eqref{eq:Qincsca_decompose} can be evaluated as:
\begin{subequations}
\begin{align}
    \matr{Q}_{qp}^{\inc,\inc} &= 0 \\
    \matr{Q}_{qp}^{\sca,\inc} &= {\cal Q}^{\sca,\inc}\left(\phi_p^\inc, \sigma_q\right) \\
    \matr{Q}_{qp}^{\inc,\sca} &= \left[\matr{Q}_{pq}^{\sca,\inc}\right]^* \label{eq:Q_incsca1}\\
    \matr{Q}_{qp}^{\sca,\sca} &= \matr{Q}_{i,qp}^{\sca,\sca} + \matr{Q}_{d,qp}^{\sca,\sca} \label{eq:Q_scasca1}\\
    \matr{Q}_{i,qp}^{\sca,\sca} &= {\cal Q}_{i}^{\sca,\sca}(\sigma_q, \sigma_p) \\
    \matr{Q}_{d,qp}^{\sca,\sca} &= {\cal Q}_{d}^{\sca,\sca}\left(\sigma_q, \sigma_p\right) 
\end{align}
\label{eq:Qdirect_Evaluation_block}
\end{subequations}
where
\begin{align}
    {\cal Q}^{\sca,\inc}\left(\phi,\sigma\right) &= \frac{1}{2k} \int_{d\Omega_s} \phi'(\vr) \sigma^*(\vr) d\vr 
\label{eq:Qscainc_expression1}
\end{align}

\begin{subequations}
\begin{align}
&{\cal Q}_{i}^{\sca,\sca}(\sigma_q,\sigma_p) \nonumber \\
&= \int_{d\Omega_s} \int_{d\Omega_s} \sigma_q^*(\vrq) \sigma_p(\vrp) \cdot \nonumber \\
&\quad \quad \quad \quad \quad \bigg[\frac{\cos (k D)}{8 \pi k^2 D} - \frac{\sin (k D)}{8 \pi k} \bigg] d\vrq d\vrp  \\
    &= \frac{1}{2k} \int_{d\Omega_s} \sigma_q^*(\vr) \mathbb{R}e \left[\frac{1}{k}{\cal L}\left[\sigma_p\right](\vr) + {\cal L}'\left[\sigma_p \right](\vr)\right] d\vr
    \end{align} \label{eq:Qi_scasca1}
\end{subequations}
\begin{subequations}
    \begin{align}
    &{\cal Q}_d^{\sca,\sca}(\sigma_q,\sigma_p) \nonumber \\
    &= \int_{d\Omega_s} \int_{d\Omega_s} \sigma_q^*(\vrq) \sigma_p(\vrp) \frac{j \unit{d} \cdot \left(\vrq + \vrp \right)}{8 \pi}  \\
    &\quad \quad \quad \quad \quad \left[ \frac{\sin(k D)}{k^2 D^2}  - \frac{\cos(k D) }{k D} \right] d\vrq d\vrp \nonumber \\
    &= \frac{j}{8k^2} \sum_{t=1}^{M} \bigg \{ \int_{d\Omega_{s}} \phi_{p}^{\inc}{'}(\vrp)  \sigma_p(\vrp) d\vrp  \nonumber \\
 & \quad \quad  \int_{d\Omega_{s}} \phi_q^{\inc*}(\vrq)  \sigma_q^*(\vrq) d\vrq -  \int_{d\Omega_{s}} \phi_p^\inc(\vrp)  \nonumber \\
 & \quad \quad \cdot \sigma_p(\vrp) d\vrp    \int_{d\Omega_{s}} \phi_q^{\text{inc}*}{'}(\vrq)  \sigma_q^*(\vrq) d\vrq
 \bigg \} 
\end{align}
\label{eq:Qd_scasca1}
\end{subequations}
In the above equations, $D = \sqrt{\left(\vrq - \vrp \right)\cdot \left(\vrq - \vrp \right)}$, $    \unit{d} = \frac{\vrq - \vrp}{D}$, $\phi_{p}^\inc{'}(\vr) = {\cal W}_p^\inc{'}(\vr)$ is the wavenumber derivative of the incident field, $\mathbb{R}e$ denotes the operation of taking the real part, and 
\begin{align}
    {\cal L}'\left[\sigma_p\right](\vr) &= \int_{d\Omega_s} \sigma_p(\vrp) G'(\vr,\vrp) d\vrp
    \label{eq:LprimeOperator}
\end{align}
is the wavenumber derivative of ${\cal L}[\sigma_p](\vr)$ under the assumption that  $\sigma_p(\vr)$ is kept constant and $G'(\vr,\vrp) = \frac{-j}{4\pi} e^{-jk\abs{\vr-\vrp}}$.\\

\noindent
\underline{Sound-Hard}:
For sound-hard scatterers, the volume integral in Eqn.~\eqref{eq:Qincsca_decompose} can be evaluated by first decomposing $\matr{Q}_{qp}$ as in Eqn.~\eqref{eq:Qincsca_decompose} and then evaluating each term using Eqn.~\eqref{eq:Qdirect_Evaluation_block}. For sound-hard scatterers, the three nonzero components of $\matr{Q}$ are
\begin{align}
    {\cal Q}^{\sca,\inc}(\phi, \sigma) = \frac{1}{2k}\int_{d\Omega_s} \frac{\partial}{\partial n} \phi'(\vr) \sigma^*(\vr) d\vr \label{eq:Qscainc_neumann}
\end{align}
\begin{subequations}
\begin{align}
&{\cal Q}_{i}^{\sca,\sca}(\sigma_q,\sigma_p) = \int_{d\Omega_s} \int_{d\Omega_s} \sigma_q^*(\vrq) \sigma_p(\vrp) \cdot \nonumber \\
&\quad \quad \frac{\partial^2}{\partial n' \partial n''}\bigg[\frac{\cos (k D)}{8 \pi k^2 D} - \frac{\sin (k D)}{8 \pi k} \bigg] d\vrq d\vrp  \\
    &= \frac{1}{2k} \int_{d\Omega_s} \sigma_q(\vr) \mathbb{R}e \left[\frac{1}{k}{\cal M}\left[\sigma_p\right](\vr) + {\cal M}'\left[\sigma_p \right](\vr)\right] d\vr
    \end{align} \label{eq:Qi_scasca1_neumann}
\end{subequations}
\begin{subequations}
    \begin{align}
    &{\cal Q}_d^{\sca,\sca}(\sigma_q,\sigma_p) \nonumber \\
    &= \int_{d\Omega_s} \int_{d\Omega_s} \sigma_q^*(\vrq) \sigma_p(\vrp) \frac{\partial^2}{\partial n' \partial n''} \bigg[\frac{j \unit{d} \cdot \left(\vrq + \vrp \right)}{8 \pi}  \\
    &\quad \quad \quad \quad \quad \left( \frac{\sin(k D)}{k^2 D^2}  - \frac{\cos(k D) }{k D} \right) \bigg] d\vrq d\vrp \nonumber \\
    &= \frac{j}{8k^2} \sum_{t=1}^{M} \bigg \{ \int_{d\Omega_{s}} \frac{\partial}{\partial n'} \phi_{p}^{\inc}{'}(\vrp)  \sigma_p(\vrp) d\vrp  \nonumber \\
 & \quad  \int_{d\Omega_{s}} \frac{\partial}{\partial n''} \phi_q^{\inc*}(\vrq)  \sigma_q^*(\vrq) d\vrq -  \int_{d\Omega_{s}} \frac{\partial}{\partial n'} \phi_p^\inc(\vrp)  \nonumber \\
 & \quad \cdot \sigma_p(\vrp) d\vrp    \int_{d\Omega_{s}} \frac{\partial}{\partial n''} \phi_q^{\text{inc}*}{'}(\vrq)  \sigma_q^*(\vrq) d\vrq
 \bigg \}\,. 
\end{align}
\label{eq:Qd_scasca1_neumann}
\end{subequations}

Eqns.~\eqref{eq:Qdirect_Evaluation_block}--\eqref{eq:Qd_scasca1_neumann} are the main result of this paper.  They show that the expensive computation of the time delay matrix $\matr{Q}$ via the volume integration of energy quantities in Eqn.~\eqref{eq:Qvolumen1} can be reduced to the evaluation of a few surface integrals presented in Eqns.~\eqref{eq:Qscainc_expression1}--\eqref{eq:Qd_scasca1} for sound-soft and Eqns.~\eqref{eq:Qscainc_neumann}--\eqref{eq:Qd_scasca1_neumann} for sound-hard scatterers. Given the reduction in dimensionality, the evaluation of these surface integrals is far more practical than that of the volume integrals in Eqn.~\eqref{eq:Qvolumen1}, and provides a practical avenue for computing the WS time delay matrix in a BEM context as demonstrated next.

\subsection{BEM Implementation}
\label{sec:direct_Q_MOM}

Implementation of the above equations within a BEM framework is straightforward.\\

\noindent
\underline{Sound-Soft}:
For sound-soft scatterers, substituting Eqns.~\eqref{eq:phi_ps_expand_dirichlet} and \eqref{eq:Vs_np} into Eqn.~\eqref{eq:Q_incsca1} yields
\begin{align}
        \matr{Q}^{\sca,\inc} =  \matJ^\dag \widetilde{\matr{Q}}^{\sca,\inc} \label{eq:discrete_Qscainc}
\end{align}
where
\begin{align}
    \widetilde{\matr{Q}}_{np}^{\sca,\inc} &= {\cal Q}^{\sca,\inc}(\phi_p^\inc, f_n) \nonumber \\
    &= -\frac{1}{2k} \matV_{np}'\,.
\end{align}
Likewise, substituting Eqn.~\eqref{eq:phi_ps_expand_dirichlet} into Eqn.~\eqref{eq:Q_scasca1} yields
\begin{align}
    \matr{Q}^{\sca,\sca} &= \matJ^\dag \left( \matr{Q}_{i}^{\sca,\sca} + \matr{Q}_d^{\sca,\sca} \right) \matJ
    \label{eq:Qscasca_discrete}
\end{align}
where
\begin{subequations}
\begin{align}
    &\matr{Q}_{i,mn}^{\sca,\sca} = {\cal Q}_i^{\sca,\sca}(f_m, f_n) \nonumber \\
    & \quad \quad \quad = \frac{1}{4k^2} \left(\matZ_{mn} + \matZ_{mn}^*\right) + \frac{1}{4k} \left(\matZ_{mn}' + \matZ_{mn}'^*\right) \nonumber \\
    & \quad \quad \quad = \frac{1}{2k} \mathbb{R}e \left(\frac{1}{k}\matZ_{mn} + \matZ_{mn}'\right) \\
    & \matr{Q}^{\sca,\sca}_{d,mn} = {\cal Q}_d^{\sca,\sca}(f_m,f_n) \nonumber \\
    & \quad \quad = \frac{j}{8k^2} \sum_{t=1}^M \matV_{mt}^* \matV_{tn}' - \left(\matV_{mt}'\right)^* \matV_{tn}\,.
\end{align}
\end{subequations}
In the above equations
\begin{subequations}
\begin{align}
     \matZ_{mn}' &= \int_{d\Omega_s} f_m(\vr) {\cal L}'\left[f_n \right](\vr) d\vr \\
 \matV_{np}' &= - \int_{d\Omega_s} \phi_p^\inc{'}(\vr) f_n(\vr) d\vr\,.
 \end{align}
 \end{subequations}
\\
Substituting Eqns.~\eqref{eq:discrete_Qscainc} and~\eqref{eq:Qscasca_discrete} into Eqn.~\eqref{eq:Qincsca_decompose} yields the following BEM expression for the WS time delay matrix:
\begin{align}
    \matr{Q} &= -\frac{1}{2k} \matJ^\dag \matV' -\frac{1}{2k} \matV'^\dag \matJ + \frac{1}{4k^2}\matJ^\dag \left(\matZ + \matZ^*\right) \matJ \nonumber \\
    &\quad + \frac{1}{4k}\matJ^\dag \left(\matZ' + \matZ^*{'}\right) \matJ \nonumber \\
    &\quad + \frac{j}{8k^2} \matJ^\dag \left(\matV^*\matV'^T - \matV^*{'}\matV\right) \matJ.
    \label{eq:Qdirect_Dirichlet}
\end{align}
In the above equation, the first two terms represent $\matr{Q}^{\sca,\inc}$ and $\matr{Q}^{\inc,\sca}$. The third and fourth terms represents contribution of $\matr{Q}_i^{\sca,\sca}$. Finally, the last term in Eqn.~\eqref{eq:Qdirect_Dirichlet} represents the contribution of $\matr{Q}_d^{\sca,\sca}$.\\

\noindent
\underline{Sound-Hard}: For sound-hard scatterers, $\matr{Q}$ can be computed using Eqns.~\eqref{eq:discrete_Qscainc}-\eqref{eq:Qdirect_Dirichlet} using the correct expressions for $\matV$ and $\matZ$ defined in Eqns.~\eqref{eq:Zmn_Neumann} and~\eqref{eq:Vh_Neumann}; 
$\matZ'$ and $\matV'$ for sound-hard scatterers are given by
\begin{subequations}
\begin{align}
     \matZ_{mn}' &= \int_{d\Omega_s} f_m(\vr) {\cal M}'\left[f_n \right](\vr) d\vr \label{eq:Discrete_Neumann_Zprime}\\
 \matV_{np}' &= - \int_{d\Omega_s} \frac{\partial}{\partial n} \phi_p^\inc{'}(\vr) f_n(\vr) d\vr\,. \label{eq:Discrete_Neumann_Vprime}
 \end{align}
 \end{subequations}

\section{Indirect Computation of $\matr{Q}$ via the Scattering Matrix and its Wavenumber Derivative}

This section presents an indirect way to compute WS time delay matrix by using the defining equation \eqref{eq:WS_relation}, i.e. as the product of $\matr{S}$ and $\matr{S}'$. Surface integral operators for computing both $\matr{S}$ and $\matr{S}'$ are presented next -- expressions for their product are lengthy and omitted.  Discretized BEM versions of the surface integral operators for $\matr{S}$ and $\matr{S}'$ are presented as well.

\subsection{Computation of the Scattering Matrix}

The scattering matrix $\matr{S}$ can be evaluated once $\phi_{p,\infty}^\sca(\vr)$ is computed near $d\Omega$. Expressions for $\matr{S}$ depend on whether $d\Omega_s$ is sound-soft or sound-hard.\\

\noindent
\underline{Sound-Soft}:
For sound-soft scatterers, $\phi_{p,\infty}^\sca(\vr)$ reads
\begin{subequations}
\begin{align}
    \phi_{p,\infty}^\sca(\vr) &=   \int_{d\Omega_s} G_{\infty}(\vr,\vrp) \sigma_p(\vrp) d\vrp \label{eq:phip_sca_infty} \\
    &= \sum_{t = 1}^{M} \matr{P}_{tp} {\cal I}_{t,\infty}^*(k r)\,
\end{align}
\end{subequations}
where $  G_\infty(\vr,\vrp) = \frac{e^{-jkr}}{4\pi r} e^{j k \unit{r} \cdot \vrp}$
and the $(t,p)$-th entry of the perturbation matrix reads
\begin{align}
    \matr{P}_{tp} &= -  \frac{j}{2 k} \int_{d\Omega_s} \phi_t^\inc(\vr) \sigma_p(\vr) d\vr\,. \label{eq:Pmat1}
\end{align}
Here, Eqns.~\eqref{eq:phip_sca_infty}--\eqref{eq:Pmat1} immediately follow from the spherical wave expansion of the Green's function in Eqn.~\eqref{eq:GSW_1}.
The total field on $d\Omega$ therefore is
\begin{align}
    &\phi_{p,\infty}(\vr) = \phi_{p,\infty}^\inc(\vr) + \phi_{p,\infty}^\sca(\vr) \nonumber \\
    & \quad = {\cal W}_{p,\infty}(k r) + \sum_{t=1}^M \matr{P}_{tp} {\cal I}_{t,\infty}^*(k r) \nonumber \\
    & \quad = {\cal I}_{p,\infty}(k r) \nonumber \\
    & \quad \quad + \sum_{t=1}^M \left((-1)^{1+l+m} \delta_{\hat{t} p} +  \matr{P}_{tp} \right) {\cal I}_{t,\infty}^*(k r) \label{eq:Pmat_Dirichilet2}
    \end{align}
where $\hat{t} = (l,-m)$. Comparing Eqn.~\eqref{eq:Pmat_Dirichilet2} to Eqn.~\eqref{eq:phitot_Dirichlet} shows that
\begin{align}
    \matr{S}_{tp} = \vec{\matr{I}}_{tp} + \matr{P}_{tp} \label{eq:Stp_cont_Dirichlet}
\end{align}
where $\vec{\matr{I}}_{tp}$ is an identity-like matrix with entries
\begin{align}
    \vec{\matr{I}}_{tp} = (-1)^{1+l+m}\delta_{\hat{t}p}\,.
\end{align}\\

\noindent
\underline{Sound-Hard}: For sound-hard scatterers, $\phi_{p,\infty}^\sca(\vr)$ reads
\begin{subequations}
\begin{align}
    \phi_{p,\infty}^\sca(\vr) &= \int_{d\Omega_s} \sigma_p(\vrp) \frac{\partial}{\partial n'} G_\infty(\vr,\vrp) d\vrp \\
    &= \sum_{t=1}^{M} \matr{P}_{tp} {\cal I}_{t,\infty}^*(kr)
\end{align}
\end{subequations}
where
\begin{align}
    \matr{P}_{tp} &= -\frac{j}{2k} \int_{d\Omega_s} \frac{\partial}{\partial n} \phi_t^\inc(\vr) \sigma_p(\vr) d\vr\,. \label{eq:Pmat_Neumann}
\end{align}
Just like sound-soft scatterers, the scattering matrix for sound-hard scatterers is also related to $\matr{P}_{tp}$ by Eqn.~\eqref{eq:Stp_cont_Dirichlet}.
Here, Eqn.~\eqref{eq:Pmat_Neumann} is obtained by taking the normal derivative of the spherical wave expansion of the Green's function in Eqn.~\eqref{eq:GSW_1}.

Eqns.~\eqref{eq:Pmat1} and \eqref{eq:Pmat_Neumann} show that the expression for the entries of perturbation matrices for the sound-soft and sound-hard cases are similar except that the normal derivative is applied to the incident field impinging a sound-hard scatterer.

\subsection{Computation of the Wavenumber Derivative of the Scattering Matrix}

Knowledge of $\sigma_p(\vr)$ also allows for the computation of wavenumber derivative of the scattering matrix. Indeed, taking the derivative of Eqns.~\eqref{eq:Pmat1} and~\eqref{eq:Pmat_Neumann}, and using $\matr{S}' = \matr{P}'$, yields the following expressions for sound-soft and sound-hard scatterers\\

\noindent
\underline{Sound-Soft}: The wavenumber derivative of Eqn.~\eqref{eq:Pmat1} is
\begin{align}
     \matr{S}_{tp}' =&  \frac{j}{2 k^2} \int_{d\Omega_s} \phi_t^\inc(\vr) \sigma_p(\vr) d\vr  \label{eq:dPmat1} \\
     & -\frac{j}{2 k} \int_{d\Omega_s} \left[\phi_t^{\inc}{'}(\vr) \sigma_p(\vr) + \phi_t^{\inc}(\vr) \sigma_p'(\vr)\right]d\vr \,. \nonumber 
\end{align}
Note that Eqn.~\eqref{eq:dPmat1} depends on $\sigma_p'(\vr)$. In order to obtain an expression for $\matr{S}_{tp}'$ that is independent of $\sigma_p'(\vr)$, consider the integral over $d\Omega_s$ of $\sigma_t(\vr)$ multiplied by the  wavenumber derivative of Eqn.~\eqref{eq:Dirichlet_IE} with $\alpha =0$. The resulting expression, simplified by invoking the symmetry of the ${\cal L}$ operator,  reads
\begin{align}
    \int_{d\Omega_s} \phi_t^\inc(\vr) \sigma_p{'}(\vr) &= \int_{d\Omega_s} \big[ \phi_p^\inc{'}(\vr) \sigma_t(\vr) \nonumber \\
    & \quad + \sigma_t(\vr) {\cal L}'\left[\sigma_p\right](\vr) \big] d\vr\,. \label{eq:DirichletPhiSigPrime1}
\end{align}
Substituting Eqn.~\eqref{eq:DirichletPhiSigPrime1} into Eqn.~\eqref{eq:dPmat1} yields
\begin{align}
     \matr{S}_{tp}' =&  \frac{j}{2 k^2} \int_{d\Omega_s} \phi_t^\inc(\vr) \sigma_p(\vr) d\vr  \label{eq:Sprime_Dirichlet1} \\
     & -\frac{j}{2 k} \int_{d\Omega_s} \big[\phi_t^{\inc}{'}(\vr) \sigma_p(\vr) + \phi_p^\inc{'}(\vr) \sigma_t(\vr) \nonumber \\
     & \quad \quad \quad \quad \quad + \sigma_t(\vr) {\cal L}'\left[\sigma_p\right](\vr) \big]d\vr \,. \nonumber
\end{align}
The entries of $\matr{Q}$ can be computed using the expressions for $\matr{S}_{tp}$ and $\matr{S}_{tp}'$ via Eqn. \eqref{eq:WS_relation}. \\

\noindent
\underline{Sound-Hard}: The wavenumber derivative of Eqn.~\eqref{eq:Pmat_Neumann} is
\begin{align}
    \matr{S}_{tp}' &= \frac{j}{2k^2} \int_{d\Omega_s} \frac{\partial}{\partial n} \phi_t^\inc(\vr) \sigma_p(\vr) d\vr  \label{eq:dPmat_Neumann} \\
    &\quad -\frac{j}{2k} \int_{d\Omega_s} \bigg[ \frac{\partial}{\partial n} \phi_t^\inc{'}(\vr) \sigma_p(\vr) \nonumber \\
    &\quad \quad + \frac{\partial}{\partial n} \phi_t^\inc(\vr) \sigma_p'(\vr) \bigg] d\vr\,. \nonumber
\end{align}
To eliminate $\sigma_p'(\vr)$ from Eqn.~\eqref{eq:dPmat_Neumann}, consider the integral over $d\Omega_s$ of $\sigma_t(\vr)$ multiplied by   the wavenumber derivative of Eqn.~\eqref{eq:cfie_neumann} with $\alpha =1$. The resulting equation reads
\begin{align}
    &\int_{d\Omega_s} \frac{\partial}{\partial n} \phi_t^\inc(\vr) \sigma_p'(\vr) d\vr \nonumber \\
    &= \int_{d\Omega_s} \sigma_t(\vr) {\cal M}'\left[\sigma_p\right](\vr) + \frac{\partial}{\partial n} \phi_p^\inc{'}(\vr) \sigma_t(\vr) d\vr\,. \label{eq:NeumannPhiSigPrime1}
\end{align}
Substituting Eqn.~\eqref{eq:NeumannPhiSigPrime1} into Eqn.~\eqref{eq:dPmat_Neumann} yields
\begin{align}
    \matr{S}'_{tp} =& \frac{j}{2k^2} \int_{d\Omega_s} \frac{\partial}{\partial n} \phi_t^\inc(\vr) \sigma_p(\vr) d\vr  \label{eq:dPmat_Neumann2} \\
     & -\frac{j}{2k} \int_{d\Omega_s} \bigg[ \frac{\partial}{\partial n} \phi_t^\inc{'}(\vr) \sigma_p(\vr)  \nonumber \\
    & + \sigma_t^d(\vr) {\cal M}'\left[\sigma_p\right](\vr) + \frac{\partial}{\partial n} \phi_p^\inc{'}(\vr) \sigma_p(\vr) \bigg] d\vr\,. \nonumber
\end{align}

\subsection{BEM Implementation}
The above equations for $\matr{S}$ and $\matr{S}'$ are easily implemented in a BEM framework for both sound-soft and sound-hard scatterers.\\

\noindent
\underline{Sound-Soft}: For sound-soft scatterers, substituting Eqns.~\eqref{eq:phi_ps_expand_dirichlet} and~\eqref{eq:Vs_np} into Eqn.~\eqref{eq:Stp_cont_Dirichlet} yields
\begin{align}
    \matr{S} = \vec{\matr{I}} + \frac{j}{2k} \matV^T \matJ \,. \label{eq:S_discrete_Dirichlet}
\end{align}
A discrete expression for $\matr{S}'$ follows from Eqn.~\eqref{eq:Sprime_Dirichlet1}
\begin{align}
    \matr{S}' &= -\frac{j}{2k^2} \matV^T \matJ + \frac{j}{2k} \matV'^T \matJ + \frac{j}{2k} \matJ^T\matV' \nonumber \\
    &\quad - \frac{j}{2k} \matJ^T \matZ' \matJ\,. \label{eq:Sprime_discrete_Dirichlet}
\end{align}
Using Eqns.~\eqref{eq:S_discrete_Dirichlet} and~\eqref{eq:Sprime_discrete_Dirichlet}, the WS time delay matrix can be expressed as
\begin{align}
    \matr{Q} &= j\left(\vec{\matr{I}} + \frac{j}{2k} \matV^T \matJ \right)^\dag \bigg(-\frac{j}{2k^2} \matV^T \matJ \nonumber \\
    & + \frac{j}{2k} \matV'^T \matJ + \frac{j}{2k} \matJ^T\matV' - \frac{j}{2k} \matJ^T \matZ' \matJ \bigg)\,.
    \label{eq:Qindirect_Discrete}
\end{align}\\
\noindent
\underline{Sound-Hard}: For sound-hard scatterers, the expressions for $\matr{S}$, $\matr{S}'$, and $\matr{Q}$ are identical to those in Eqns.~\eqref{eq:S_discrete_Dirichlet}, \eqref{eq:Sprime_discrete_Dirichlet}, and~\eqref{eq:Qindirect_Discrete} except the $\matV$, $\matZ$, $\matV'$, and $\matZ'$ are defined as in Eqns.~\eqref{eq:Vh_Neumann}, \eqref{eq:Zmn_Neumann}, \eqref{eq:Discrete_Neumann_Vprime}, \eqref{eq:Discrete_Neumann_Zprime}, respectively.\\

Though it is not immediately apparent that the indirect approach presented in this section yields identical results to the direct approach, the equivalence of both approaches is demonstrated in Appendix~\ref{app:direct_indirect_equivalence}.

\section{Numerical Examples}

This section demonstrates the application of the above BEM framework for computing the WS time delay matrix for two 3D geometries: two parallel plates and a pill-box with a slot. For brevity, only results obtained using the ``indirect'' approach are presented. For both examples, the scatterers are sound-soft, and the wave speed in the exterior medium is $v=1$ m/s. A BEM code using pulse basis functions defined on a triangular patch surface model of the scatterer is used to construct $\matr{Q}$.


\subsection{Parallel Plate}

\begin{figure*}[hbt!]
\figline{\fig{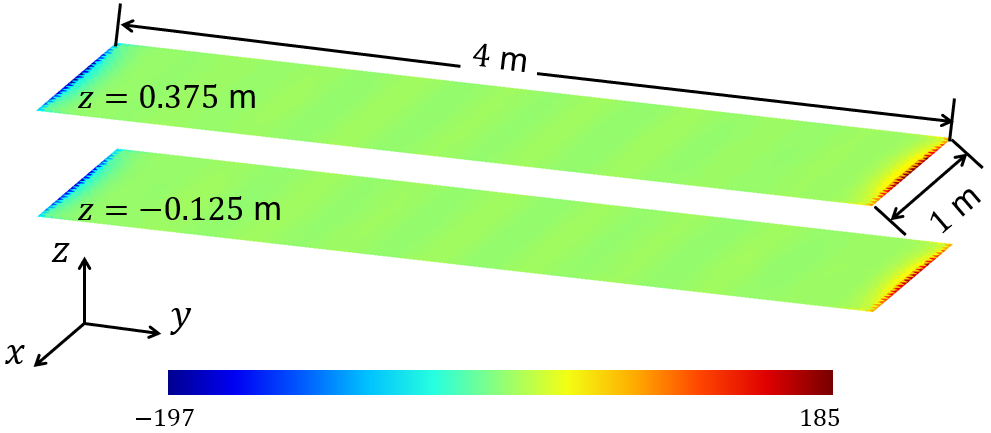}{0.3\textwidth}{(a) WS mode \#1}\label{fig:para_plate_WS1} 
\fig{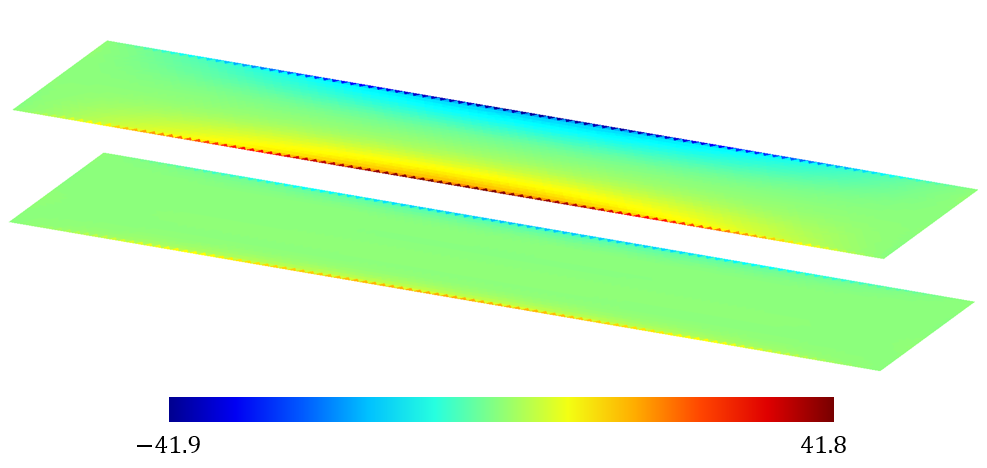}{0.3\textwidth}{(b) WS mode \#24}\label{fig:para_plate_WS24}
\fig{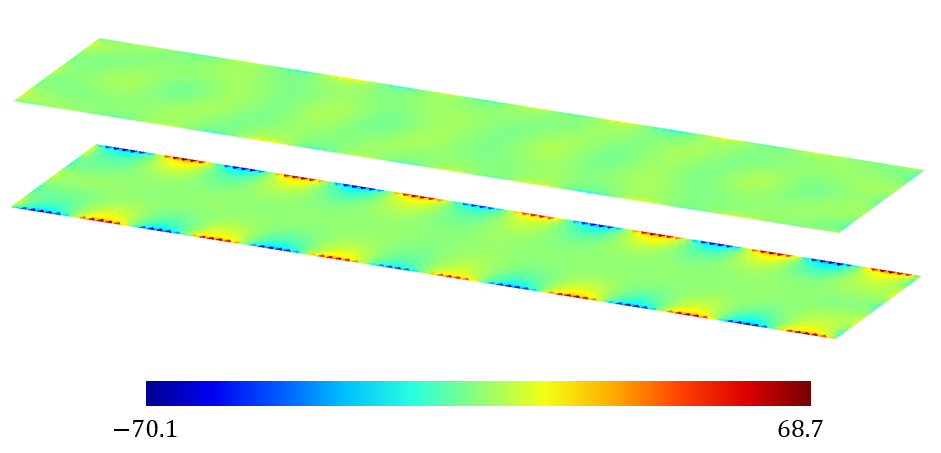}{0.3\textwidth}{(c) WS mode \#120}\label{fig:para_plate_WS120}}
\figline{\fig{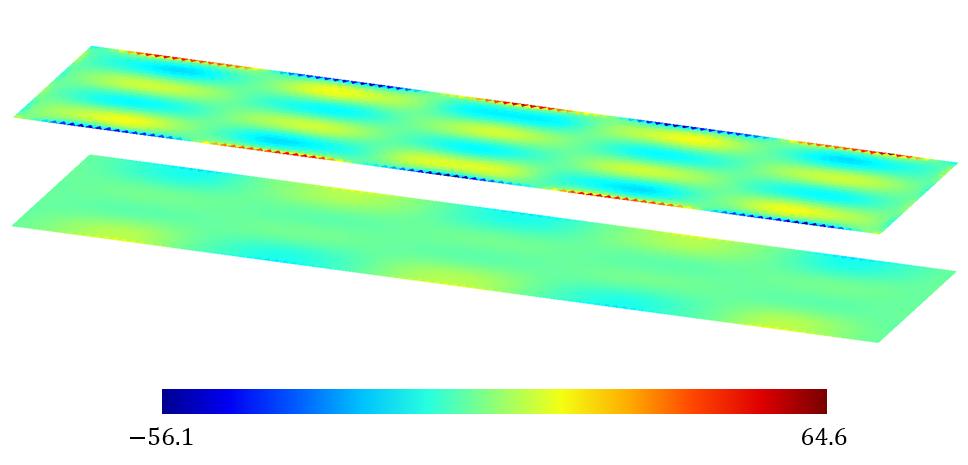}{0.3\textwidth}{(d) WS mode \#140}\label{fig:para_plate_WS140} 
\fig{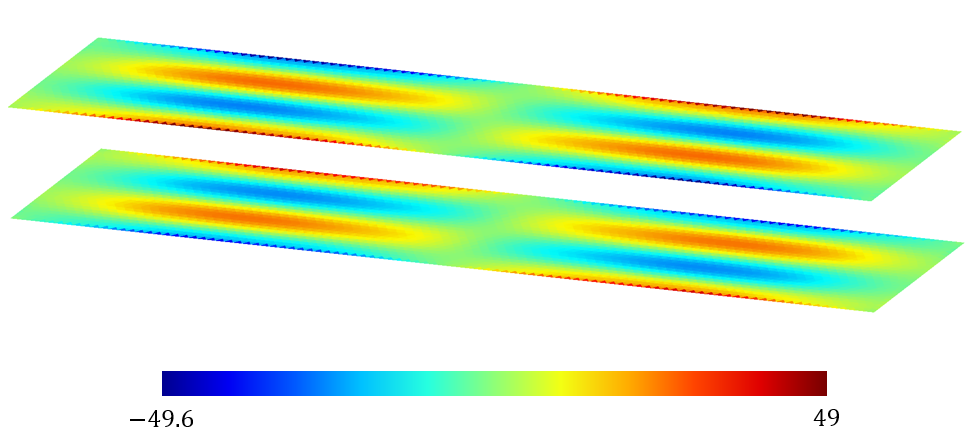}{0.3\textwidth}{(e) WS mode \#1280}\label{fig:para_plate_WS1280}
\fig{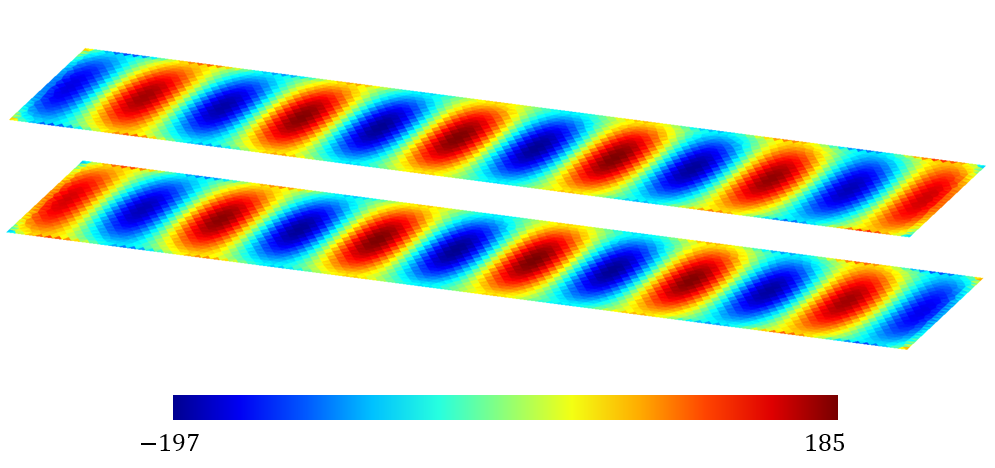}{0.3\textwidth}{(f) WS mode \#1296}\label{fig:para_plate_WS1296}}
\caption{Selected WS modes (real part) of the sound-soft parallel plate scatterer.}
\label{fig:WSmodes_para_plate}
\end{figure*}

A sound-soft parallel plate scatterer (Fig.~\ref{fig:WSmodes_para_plate}a) is studied first. The scatterer is embedded in a medium with $k=15.708$ m$^{-1}$ and is illuminated by $M=1296$ incoming spherical harmonics. 



\begin{figure}[t]
\includegraphics{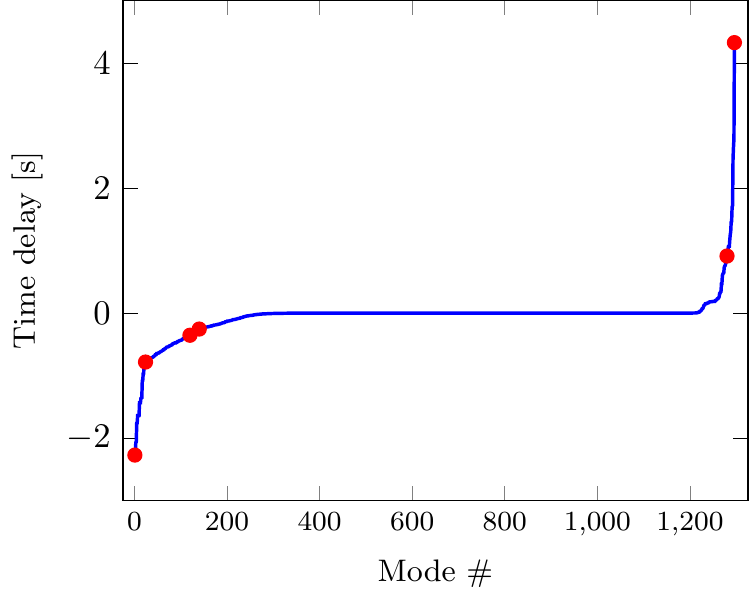}
\caption{Time delays of the WS modes of parallel plate. Red dots represent selected WS modes in Fig.~\ref{fig:WSmodes_para_plate}.}
\label{fig:para_plate_time_delay}
\end{figure}


The matrix $\matr{Q}$ is diagonalized as $\matr{Q} = \matr{W} \overline{\matr{Q}}\matr{W}^\dag$ and $\overline{\matr{Q}}$'s diagonal elements, viz. the WS time delays or eigenvalues of $\matr{Q}$, are shown in Fig.~\ref{fig:para_plate_time_delay}. Note that the time delays can be converted to spatial distances through multiplication by $v$. 
The surface source densities for the WS modes, i.e. $\sigma_{WS,q} = \sum_p \matr{W}_{pq} \sigma_p$, are analysed next. 
In what follows, WS modes are ordered by their time-delays (small to large) and indexed accordingly.
The WS modes for this scatterer can be categorized into several groups:
\begin{enumerate}[a., leftmargin=* ]
    \item \emph{Short-edge modes}: The first several WS modes scatter off the four short edges and never reach the origin; therefore their corresponding time delays are negative and smaller than those of all other modes. The surface source density of mode 1 is shown in Fig.~\ref{fig:WSmodes_para_plate}a.
    
    \item \emph{Long-edge modes}: The WS modes in this group are very similar to those in the first group, except that they concentrate along the long edges of the plates, which can support a much larger number of modes associated with different incident directions compared to the short edges.
    Mode 24 (Fig.~\ref{fig:WSmodes_para_plate}b) excites the long edges nearly head-on, while mode 120 (Fig.~\ref{fig:WSmodes_para_plate}c) excites the long edges at a near-grazing angle. The time delays of the long-edge modes are smaller than those of the short-edge modes since the long edges reside closer to the origin.
    
    \item \emph{Surface ballistic modes}: WS modes can also specularly reflect off either the top or bottom plate and excite the whole surface of the plate instead of only its edges. Mode 140 (Fig.~\ref{fig:WSmodes_para_plate}d) involves a beam-like excitation of the top plate. These modes also have negative time delays, as their fields do not reach the origin which is located between the two plates.
    
    \item \emph{Non-propagating modes}: These modes do not excite both plates and experience roughly zero time delays. 
    
    \item \emph{Resonant modes}: These modes involve fields that get (weakly) trapped between the two plates before escaping. Their dwell times depend on their angle of arrival. The incident waves for mode 1280 and mode 1296 (Figs.~\ref{fig:WSmodes_para_plate}e and \ref{fig:WSmodes_para_plate}f) reflect multiple times between the two plates and are trapped for a period of time that depends on their incident angles. The corresponding time delays are positive. 
\end{enumerate}

\subsection{Pill-shaped Cavity}


\begin{figure}[t]
\includegraphics{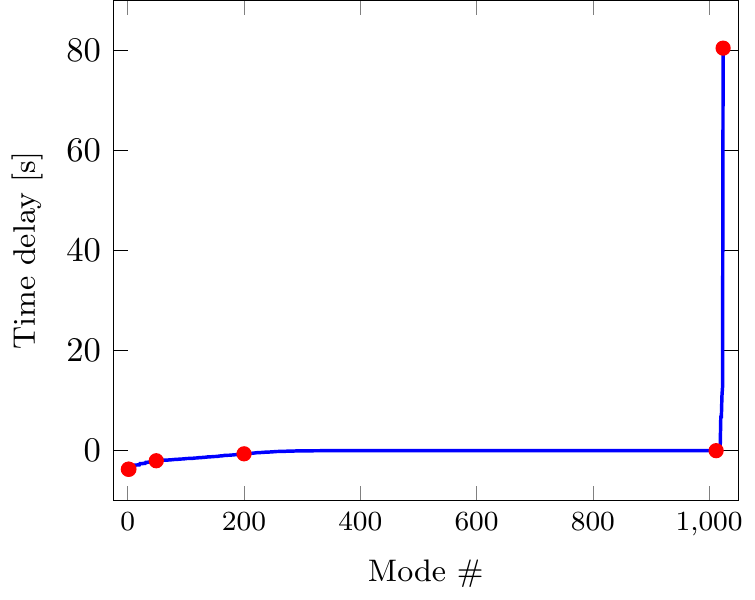}
\caption{Time delays of the pill-shaped cavity. Red dots represent selected WS modes in Fig.~\ref{fig:WSmodes_pill}.}
\label{fig:pill_time_delay}
\end{figure}

\begin{figure*}[hbt!]
\figline{\fig{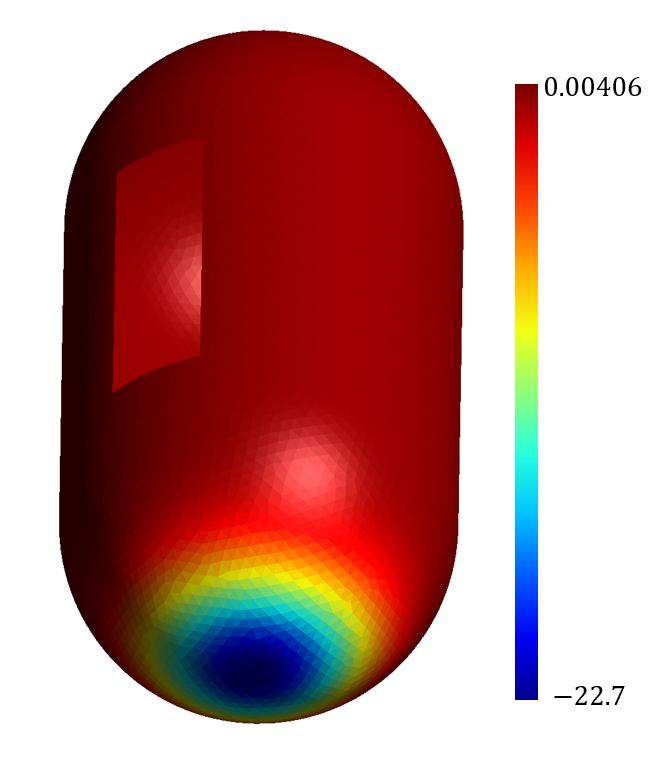}{0.3\textwidth}{(a) WS mode \#1}\label{fig:pill_WS1} 
\fig{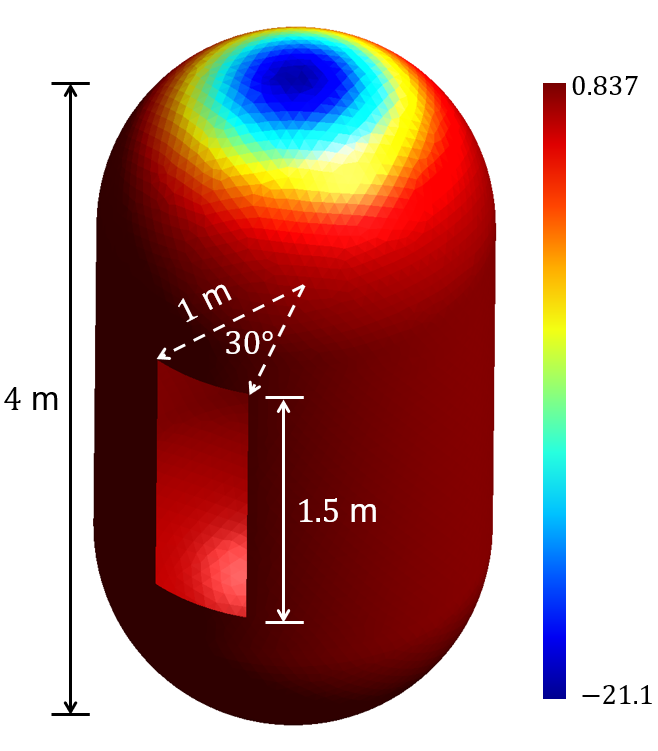}{0.3\textwidth}{(b) WS mode \#2}\label{fig:pill_WS2}
\fig{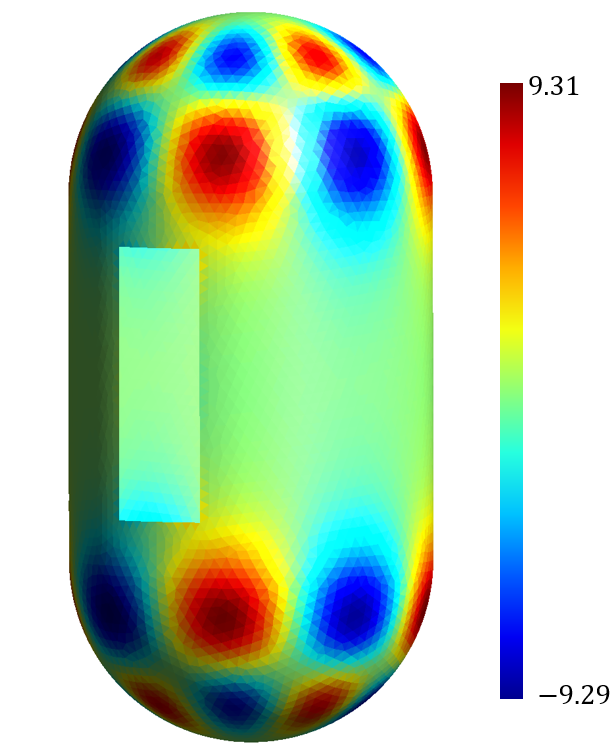}{0.3\textwidth}{(c) WS mode \#49}\label{fig:pill_WS49}}
\figline{\fig{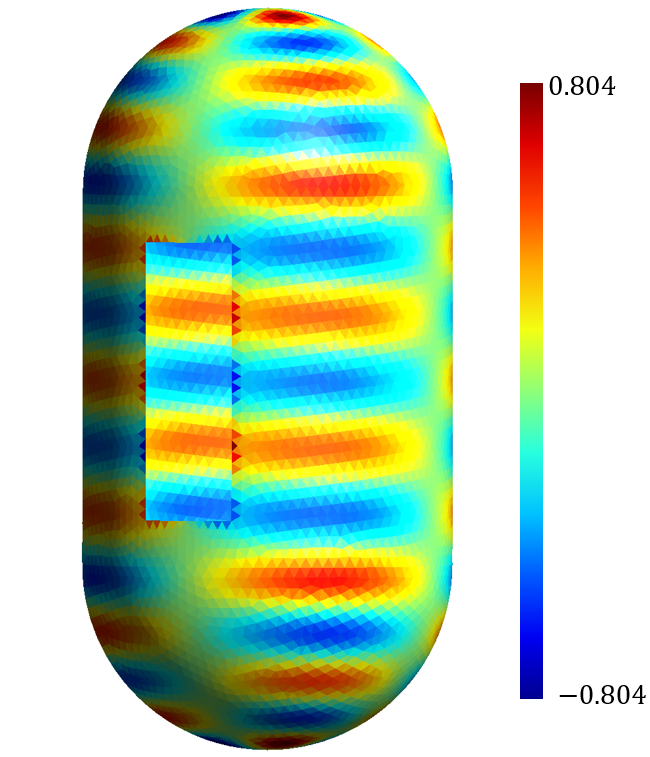}{0.3\textwidth}{(d) WS mode \#200}\label{fig:pill_WS200} 
\fig{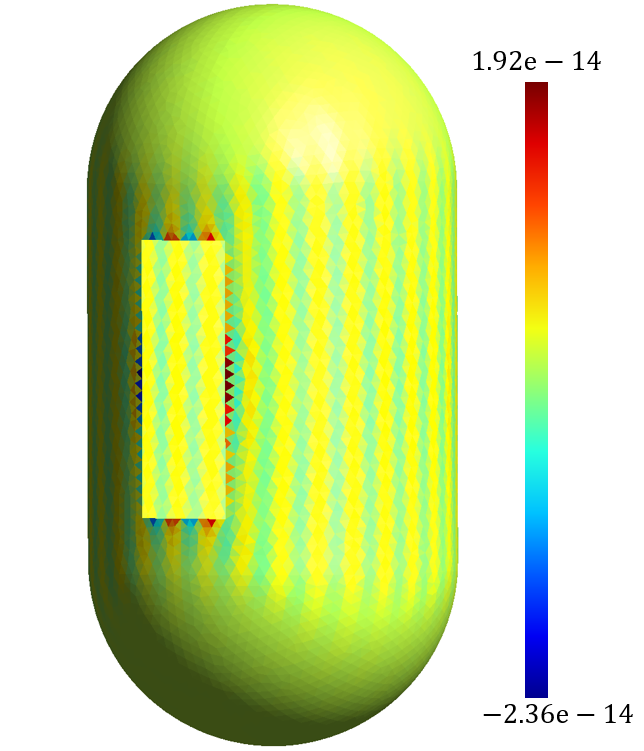}{0.3\textwidth}{(e) WS mode \#1012}\label{fig:pill_WS1012}
\fig{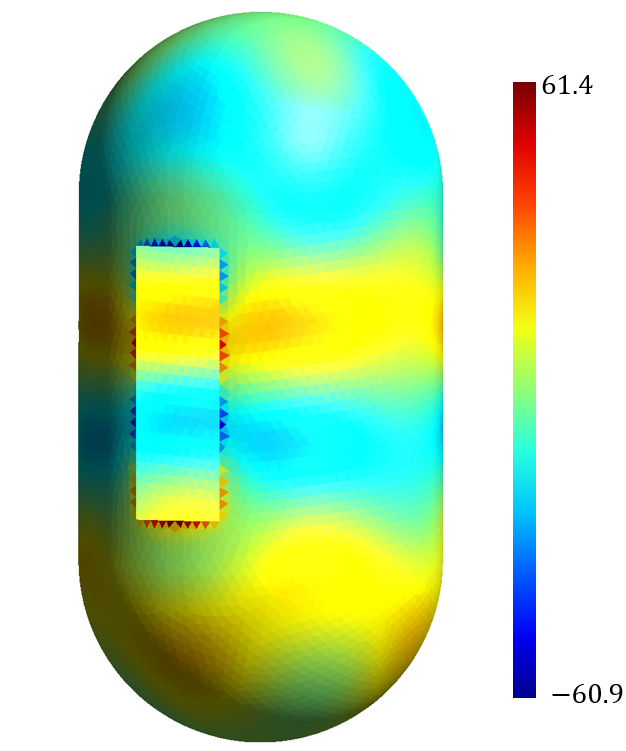}{0.3\textwidth}{(f) WS mode \#1024}\label{fig:pill_WS1024}}
\caption{Selected WS modes (real part) of the sound-soft pill-shaped cavity.}
\label{fig:WSmodes_pill}
\end{figure*}

A sound-soft pill-shaped cavity with rectangular slot (Fig.~\ref{fig:WSmodes_pill}b), embedded in a medium with $k=9.664$ m$^{-1}$, is illuminated by $M = 1024$ incoming spherical harmonics. The diagonalization of $\matr{Q}$ obtained using Eqn.~\eqref{eq:Qdirect_Dirichlet} yields WS time delays and WS modes shown in Figs.~\ref{fig:pill_time_delay} and \ref{fig:WSmodes_pill}, respectively. 
The WS modes can be classified into the following groups:
\begin{enumerate}[a., leftmargin=* ]
    \item \emph{Exterior modes}: 
    These modes scatter off the cavity's exterior, and hence experience negative time delays as they never reach the origin located inside the scatterer.
    In contrast to the parallel plate, the exterior of the pill-shaped cavity is smooth; hence the WS modes do not decouple into edge and ballistic types. Modes 1 and 2 experience negative time delays of -3.714 s and are incident from the vertical direction, exciting the top and bottom of the pill (Figs.~\ref{fig:WSmodes_pill}a and \ref{fig:WSmodes_pill}b). 
    As the incident direction gets closer to the horizontal plane, WS modes increasingly excite the lateral boundary of the pill, while the magnitude of negative time delay decreases. Mode 200 primarily excites the lateral exterior boundary (Fig.~\ref{fig:WSmodes_pill}d) and experiences a time delay of -0.631 s. The exterior modes avoid exciting the interior of cavity. 
    
    \item \emph{Non-propagating modes}: WS modes that do not excite the scatterer have roughly zero time delays. Mode 1012 has zero time delay and the surface source density is also nearly zero (Fig.~\ref{fig:WSmodes_pill}e), indicating that the scatterer is not excited.
    
    \item \emph{Interior resonant modes}: For the cavity structure, incoming waves can enter the aperture and get trapped in the cavity for a while before exiting. Mode 1024 (Fig.~\ref{fig:WSmodes_pill}f) has the largest positive time delay, 80.506 s, which corresponds to ten round trips between the top and bottom ends inside the cavity, indicating a strongly resonant behavior. 
\end{enumerate}

Both the above examples elucidate the physical meaning of the WS time delay matrix and its usefulness in untangling canonical scattering phenomena in 3D acoustic surface scattering phenomena.  

\section{Conclusion}

Two methods for computing the WS time delay matrix $\matr{Q}$ associated with acoustic scattering phenomena were described. The direct formulation computes the entries of $\matr{Q}$ by reducing the volume integral of the renormalized energy density of the field surrounding the scatterer to a surface integral. The indirect method proceeds by multiplying the scattering matrix and its frequency derivative. Both techniques are easily implemented in a BEM framework.  Knowledge of $\matr{Q}$ can be used to compute $\matr{S}'$ and wavenumber derivatives of other port variables. Diagonalization of $\matr{Q}$ reveals WS modes that interact with the scatterer with well-defined group delays.

\appendix

\section{Modes for Scattering Systems}
\label{Appdix:modes}

\subsection{Spherical Waves}
Helmholtz fields inside a spherical shell in $\mathbb{R}^3$ can be expressed as
\begin{align}
    \phi(\vr) &= \sum_{p} a_p {\cal B}_p(\vr)
\end{align}
where ${\cal B}_p(\vr) = {\cal I}_p(\vr)$, ${\cal O}_p(\vr)$, or ${\cal W}_p(\vr)$ representing incoming, outgoing, or standing waves.
The incoming waves for $p = (l,m)$ are defined as
\begin{align}
    {\cal I}_{p}(\vr) &= k j^{l+1} h_l^{(1)}(kr) {\cal X}_{lm}(\theta,\phi)
    \label{eq:Ip_def}
\end{align}
where $p=(l,m)$ for $l=0,...,L$ and $m=-l,...,l$,  $h_l^{(1)}(z)$ is the spherical Hankel function of type 1 and order $l$, 
and ${\cal X}_{p}(\theta,\phi)$ is the spherical harmonic defined as
\begin{align}
    {\cal X}_{p}(\theta,\phi) = (-1)^m \sqrt{\frac{2l+1}{4 \pi} \frac{(l-m)!}{(l+m)!}} P_l^m(\cos \theta) e^{j m \phi}.
\label{eq:Ylm}
\end{align}
Here, $P_{l}^{m}(x)$ is the associated Legendre polynomial of degree $l$ and order $m$~\citep{Abr64}. 
Outgoing spherical waves are obtained by swapping the $h_l^{(1)}(kr)$ in Eqn.~\eqref{eq:Ip_def} for $h_l^{(2)}(kr)$, the spherical Hankel function of type $2$.
Likewise, standing waves are obtained by swapping $h_l^{(1)}(kr)$ in Eqn.~\eqref{eq:Ip_def} for $2 j_l(kr)$, the regular spherical Bessel function.
The outgoing and standing waves can be expressed in terms of incoming spherical waves as
\begin{subequations}
\begin{align}
    {\cal O}_{lm}(\vr) &= (-1)^{1 + l+m} {\cal I}_{l(-m)}^*(\vr) \\
    {\cal W}_{lm}(\vr) &= {\cal I}_{lm}(\vr) + {\cal O}_{lm}(\vr) \\
    &= {\cal I}_{lm}(\vr) + (-1)^{1 + l+m} {\cal I}_{l(-m)}^*(\vr)\,.
\end{align}
\end{subequations}

\subsection{Properties of Spherical Waves}

In addition to the key orthonormality property in Eqn.~\eqref{eq:X_p_orthonorm}, the spherical harmonics satisfy other important properties used in the derivations in this paper.

\begin{enumerate}[leftmargin=*]
\item Conjugation: Spherical harmonics are related to their conjugate as
\begin{align}
    {\cal X}_{p}^*(\theta,\phi) &= (-1)^m {\cal X}_{\tilde{p}} (\theta,\phi)
\end{align}
where $\tilde{p} = (l,-m)$.
\item Large-argument approximation: as $r\rightarrow \infty$, the incoming, outgoing, and standing waves (${\cal I}_{p}(\vr)$, ${\cal O}_{p}(\vr)$, and ${\cal W}_{p}(\vr)$) can be approximated by
\begin{subequations}
\begin{align}
    {\cal I}_{p,\infty}(\vr) &\cong \frac{e^{jkr}}{r} {\cal X}_{lm}(\theta,\phi) \\
    {\cal O}_{p,\infty}(\vr) &\cong (-1)^{l+1} \frac{e^{-jkr}}{r} {\cal X}_{lm}(\theta,\phi)\\
    {\cal W}_{p,\infty}(\vr) &\cong \left[\frac{e^{jkr}}{r} + (-1)^{l+1}\frac{e^{-jkr}}{r} \right] {\cal X}_{lm}(\theta,\phi)
\end{align}
\end{subequations}
by using the large argument approximation of $h_l^{(1)}(z)$, $h_l^{(2)}(z)$, and $j_l(z)$~\citep{Kris2014VSW}.
\item Spherical wave expansion of $G(\vr,\vrp)$: For $\abs{\vr} > \abs{\vrp}$, the 3D Green's function of a homogeneous medium reads
\begin{subequations}
\begin{align}
    G(\vr,\vrp) &= \frac{1}{4\pi \abs{\vr-\vrp}} e^{-jk\abs{\vr - \vrp}} \nonumber \\
    &= -\frac{j}{2k} \sum_{p} {\cal W}_p(\vrp) {\cal I}_p^*(\vr)     \label{eq:GSW_1}
 \\
    &= -\frac{j}{2k} \sum_{p} {\cal W}_p^*(\vrp) {\cal O}_p(\vr)\,. \label{eq:GSW_2}
\end{align}
\end{subequations}
Here, the summation is performed over all possible spherical harmonics, i.e. $l = 0,\hdots, \infty$ and $m=-l,\hdots,l$.
For $r\rightarrow \infty$, the Green's function can be further simplified to read
\begin{align}
    G_\infty(\vr,\vrp) &= \frac{e^{-jk\abs{\vr}} e^{jk\vrp \cdot \unit{r}}}{4\pi \abs{\vr}} \nonumber\\
    &= \frac{-j}{2k} \sum_{m} {\cal W}_m(\vrp) {\cal I}_{m,\infty}^*(\vr)\,.
    \label{eq:GSWinf_1}
\end{align}
\end{enumerate}

\section{Direct \& Indirect form. Equivalence}
\label{app:direct_indirect_equivalence}
\subsection{Identities Obtained Analytically}
\label{app:analytic_identity}
This section proves $\matZ - \matZ^* = -\frac{j}{2k} \matV^* \matV^T$ using the Spherical wave expansion of the Green's function.\\

\noindent 
\underline{Sound-Soft}: 
Using Eqns.~(A7a) and~(A7b) in the original manuscript, for $\abs{\vrp} \le \abs{\vr}$
\begin{align}
    \matZ_{mn} &= -\frac{j}{2k} \sum_{t=1}^{\infty} \int_{d\Omega_s} f_m(\vr) {\cal I}_t^*(\vr) \cdot \nonumber \\
    &\quad \quad \int_{d\Omega_s} f_n(\vrp) {\cal W}_t(\vrp) d\vrp d\vr \label{eq:Zmn_expand1} \\
    &= -\frac{j}{2k} \sum_{t=1}^{\infty} \int_{d\Omega_s} f_m(\vr) {\cal O}_t(\vr) \cdot \nonumber \\
    &\quad \quad \int_{d\Omega_s} f_n(\vrp) {\cal W}_t^*(\vrp) d\vrp d\vr \label{eq:Zmn_expand2} \,.
\end{align}
Using Eqns.~\eqref{eq:Zmn_expand1} and \eqref{eq:Zmn_expand2}, 
\begin{align}
    \matZ_{mn} - \matZ_{mn}^* &= -\frac{j}{2k} \sum_{t=1}^{\infty} \int_{d\Omega_s} f_m(\vr) \partial {\cal W}_t^*(\vr) \cdot \nonumber \\
    &\quad \quad \int_{d\Omega_s} f_n(\vrp) {\cal W}_t(\vrp) d\vr d\vrp\,.
\end{align}
The above can be compactly expressed as the $(m,n)$-th entry of $-\frac{j}{2k} \matV^*\matV^T$. The case of $\abs{\vr} \le \abs{\vrp}$ follows from the symmetry of $\matZ$.\\

\noindent
\underline{Sound-Hard}:
Using Eqns.~(A7a) and~(A7b) in the original manuscript, for $\abs{\vrp} \le \abs{\vr}$
\begin{align}
    \matZ_{mn} &= -\frac{j}{2k} \sum_{t=1}^{\infty} \int_{d\Omega_s} f_m(\vr) \frac{\partial}{\partial n} {\cal I}_t^*(\vr) \cdot \nonumber \\
    &\quad \quad \int_{d\Omega_s}  f_n(\vrp) \frac{\partial}{\partial n'} {\cal W}_t(\vrp) d\vrp d\vr \label{eq:Zmn_expand1_neumann} \\
    &= -\frac{j}{2k} \sum_{t=1}^{\infty} \int_{d\Omega_s} f_m(\vr) \frac{\partial}{\partial n} {\cal O}_t(\vr) \cdot \nonumber \\
    &\quad \quad \int_{d\Omega_s}  f_n(\vrp) \frac{\partial}{\partial n'} {\cal W}_t^*(\vrp) d\vrp d\vr \label{eq:Zmn_expand2_neumann} \,.
\end{align}
Using Eqns.~\eqref{eq:Zmn_expand1_neumann} and \eqref{eq:Zmn_expand2_neumann} yields $(m,n)$-th entry of $-\frac{j}{2k} \matV^*\matV^T$.

\subsection{Proof}

This section shows the equivalence between the expressions for $\matr{Q}$ obtained with the direct approach in Eqn.~(36) of the original manuscript and the indirect approach in Eqn.~(53) of the original manuscript.
Using $\vec{\matr{I}}^\dag = \vec{\matr{I}}$, $\vec{\matr{I}}^T \matV^T  = \matV^\dag$, and  $\matJ' = \matZ^{-1} \left(\matV' - \matZ' \matJ \right)$ into Eqn.~(53) of the original manuscript yields  
\begin{align}
    \matr{Q} =&  -\frac{1}{2k} \matJ^\dag \matV'- \frac{1}{2k} \matV'^\dag \matJ \nonumber \\
    & \quad + \frac{1}{2k} \matJ^\dag  \left( \matZ' + \frac{1}{k} \matZ^* \right) \matJ \nonumber \\
    &\quad  - \frac{j}{4k^3} \matJ^\dag \matV^*\matV^T \matJ + \frac{j}{4k^2} \matJ^\dag \matV^*\matV'^T \matJ   \nonumber \\
&\quad       + \frac{1}{2k} \matJ^\dag \left( \matZ - \matZ^* + \frac{j}{2k} \matV^* \matV^T \right) \matJ'\,. 
\end{align}
Using the $\matZ - \matZ^* = -\frac{j}{2k} \matV^* \matV^T$ (proof in Appendix~\ref{app:analytic_identity}) simplifies the above equation to
\begin{align}
    \matr{Q} =&  -\frac{1}{2k} \matJ^\dag \matV'- \frac{1}{2k} \matV'^\dag \matJ \nonumber \\
    &\quad + \frac{1}{2k} \matJ^\dag  \left( \matZ' + \frac{1}{k} \matZ^* \right) \matJ \nonumber \\
    &\quad  - \frac{j}{4k^3} \matJ^\dag \matV^*\matV^T \matJ + \frac{j}{4k^2} \matJ^\dag \matV^*\matV'^T \matJ\,. 
\end{align}
Finally, the Eqn.~(36) of the original manuscript is obtained by using the self-adjoint property $\matr{Q} = \frac{1}{2}\left( \matr{Q} + \matr{Q}^\dag \right)$.

\section{Direct computation of Q}
\label{app:direct_comp_of_Q}
\subsection{Spherical wave expressions}

\noindent
\underline{Incident field}: Through expansion by the spherical wave functions, the incident field with mode index $p$ is written as
\begin{align}
    \phi_p^{inc} &= 2k j^{l+1} j_l(kr) Y_{lm}(\theta, \varphi)
\end{align}
where $p$ maps to a tuple $(l,m)$. When $kr \rightarrow \infty$, the approximation for $\phi_p^{inc}$ is
\begin{align}
\label{eq:phi_inc_approx_start}
    \phi_{p,\infty}^{inc} &= \left[ \frac{e^{jkr}}{r} + (-1)^{(l+1)}                \frac{e^{-jkr}}{r} \right] Y_{lm}(\theta, \varphi) \\
    \phi_{p,\infty}^{inc}{'} &= j\left[ e^{jkr} + (-1)^l e^{-jkr} \right] Y_{lm}(\theta, \varphi) \\
    \hat{\vr} \cdot \nabla \phi_{p,\infty}^{inc} &= \Big[ (jk-\frac{1}{r})\frac{e^{jkr}}{r} \nonumber \\
    & \quad + (-1)^l (jk+\frac{1}{r})\frac{e^{-jkr}}{r} \Big] Y_{lm}(\theta, \varphi) \\
    \hat{\vr} \cdot \nabla \phi_{p,\infty}^{inc}{'} &= -k\left[ e^{jkr} + (-1)^{(l+1)}e^{-jkr} \right] Y_{lm}(\theta, \varphi) \,.
    \label{eq:phi_inc_approx_end}
\end{align}

\noindent
\underline{Scattered field}: The scattered field can be expressed through the scattering parameters as
\begin{align}
    \phi_p^{sca} &= \sum_t \matr{P}_{tp} k (-j)^{(l+1)} h_{l'}^{(2)}(kr) Y_{l'm'}^*(\theta, \varphi )
\end{align}
where $t$ maps to a tuple $(l',m')$.
When $kr \rightarrow \infty$, the approximation for $\phi_p^{sca}$ is
\begin{align}
\label{eq:phi_sca_approx_start}
    \phi_{p,\infty}^{sca} &= \sum_t \matr{P}_{tp} \frac{e^{-jkr}}{r} Y_{l'm'}^*(\theta, \varphi ) \\
    \phi_{p,\infty}^{sca*} &= \sum_t \matr{P}_{tp}^* \frac{e^{jkr}}{r} Y_{l'm'}(\theta, \varphi )\\
    \hat{\vr} \cdot \nabla \phi_{p,\infty}^{sca*} &= \sum_t \matr{P}_{tp}^* (jk-\frac{1}{r}) \frac{e^{jkr}}{r} Y_{l'm'}(\theta, \varphi ) \,.
    \label{eq:phi_sca_approx_end}
\end{align}
Alternatively, the scattered field can be expressed by the integration of Green's function,
\begin{align}
    \phi_p^{sca} &= \int_{d\Omega_s} G(\vr,\vrp) \sigma_p(\vrp) dS' \nonumber \\
    \phi_{p,\infty}^{sca} &= \int_{d\Omega_s} G_\infty(\vr,\vrp) \sigma_p(\vrp) dS' \nonumber \\
    \phi_{p,\infty}^{sca}{'} &= \int_{d\Omega_s} \Big[ G'_\infty(\vr,\vrp) \sigma_p(\vrp)  \nonumber \\
    &\quad + G_\infty(\vr,\vrp) \sigma'_p(\vrp) \Big] dS'
\end{align}
where the Green's function $G$ and the far-field approximation $G_\infty$ read
\begin{align}
    G(\vr,\vrp) &= \frac{e^{-jk|\vr-\vrp|}}{4 \pi |\vr-\vrp|} \nonumber \\
    G_\infty(\vr,\vrp) &= \frac{e^{-jk(r - \hat{\vr} \cdot \vrp)}}{4 \pi r} \nonumber \\
    G_\infty'(\vr,\vrp) &= \frac{-j}{4 \pi} e^{-jk(r - \hat{\vr} \cdot \vrp)} \nonumber \\
    &\quad + \frac{j \hat{r}\cdot \vrp}{4 \pi r} e^{-jk(r - \hat{\vr} \cdot \vrp)} \nonumber \\
    \hat{\vr} \cdot \nabla G_\infty(\vr,\vrp) &= -\frac{jk}{4 \pi r} e^{-jk(r-\hat{\vr}\cdot\vrp)} \nonumber \\
    \hat{\vr} \cdot \nabla G_\infty'(\vr,\vrp) &= \frac{e^{-jk(r-\hat{\vr}\cdot\vrp)}}{4 \pi r} ( - j - kr + k \hat{\vr}\cdot\vrp) \nonumber \,.
\end{align}

\subsection{Conversion of $\matr{Q}_{qp}^{\alpha,\beta}$ to a computationally feasible form}

Consider two sets of scalar fields: $\phi_p^\beta(\vr)$ and $\phi_q^\alpha(\vr)$, $\alpha,\beta \in \{inc,sca\}$. They satisfy the following Helmholtz equation for $\vr \in \Omega$
\begin{subequations}
\begin{align}
    \nabla^2 \phi_p^\beta + k^2 \phi_p^\beta &= -\sigma_p \label{eqap:Helm1} \\
    \nabla^2 \phi_q^\alpha + k^2 \phi_q^\alpha &= -\sigma_q \label{eqap:Helm2}
\end{align}
\end{subequations}
where $k$ is the wavenumber inside $\Omega$. Taking the derivative of Eq.~\eqref{eqap:Helm1} w.r.t. $k$ and taking the conjugate of Eq.~\eqref{eqap:Helm2} yields
\begin{subequations}
\begin{align}
    \nabla^2 \phi_p^\beta{'} + 2k  \phi_p^\beta + k^2 \phi_p^\beta{'} &= -\sigma_p' \label{eqap:Helm_dk} \\
    \nabla^2 \phi_q^{\alpha*} + k^2 \phi_q^{\alpha*} &= -\sigma_q^* \,. \label{eqap:Helmconj}
\end{align}
\end{subequations}
Subtracting the product of Eq.~\eqref{eqap:Helm_dk} and $\phi_q^{\alpha*}$ from the product of Eq.~\eqref{eqap:Helmconj} and $\phi_p^\beta{'}$ yields
\begin{align}
    \phi_p^\beta{'} \nabla^2 \phi_q^{\alpha*} &- \phi_q^{\alpha*} \nabla^2 \phi_p^\beta{'} 
    + \phi_p^\beta{'} \sigma_q^* - \sigma_p' \phi_q^{\alpha*} 
    &= 2 k \phi_p^\beta \phi_q^{\alpha*}\,. \label{eqap:2a}
\end{align}
Using identity $\psi \nabla^2 \gamma - \gamma \nabla^2 \psi = \nabla \cdot \left(\psi \nabla \gamma - \gamma \nabla \psi \right)$, Eq.~\eqref{eqap:2a} simplifies to
\begin{align}
    &\frac{1}{2k} \nabla \cdot \left( \phi_p^\beta{'} \nabla \phi_q^{\alpha*} - \phi_q^{\alpha*} \nabla \phi_p^\beta{'} \right) \nonumber \\
    &\quad \quad + \frac{1}{2k} (\phi_p^\beta{'} \sigma_q^* - \sigma_p' \phi_q^{\alpha*}) 
    = \frac{1}{2}\phi_p^\beta \phi_q^{\alpha*} + \frac{1}{2}\phi_p^\beta \phi_q^{\alpha*} \label{eqap:3a}
\end{align}
The RHS of Eq.~\eqref{eqap:3a} is split on purpose. Write one of the terms,  $\frac{1}{2}\phi_p^\beta \phi_q^{\alpha*}$, as
\begin{align}
    \frac{1}{2} \phi_p^\beta \phi_q^{\alpha*}
    &= - \frac{1}{2 k^2} \phi_p^\beta \nabla^2 \phi_q^{\alpha*}-\frac{1}{2 k^2} \phi_p^\beta \sigma_q^* \nonumber \\
    &= - \frac{1}{2 k^2} \nabla \cdot (\phi_p^\beta  \nabla \phi_q^{\alpha*} )
    + \frac{1}{2 k^2} \nabla \phi_q^{\alpha*}  \cdot \nabla \phi_p^\beta \nonumber \\
    & \quad - \frac{1}{2 k^2} \phi_p^\beta \sigma_q^* \,.
    \label{eqap:4a}
\end{align}
where Eq.~\eqref{eqap:Helmconj} and the relationship $\phi \nabla^2 \gamma = \nabla \cdot (\phi \nabla \gamma) - \nabla \gamma \cdot \nabla \phi$ are made use of.
Substituting Eq.~\eqref{eqap:4a} into Eq.~\eqref{eqap:3a} yields
\begin{align}
    \frac{1}{2k} &\nabla \cdot \big(\phi_p^\beta{'} \nabla \phi_q^{\alpha*}  - \phi_q^{\alpha*} \nabla \phi_p^\beta{'}
    + \frac{1}{k} \phi_p^\beta \nabla \phi_q^{\alpha*} \big) \nonumber \\
    &+\frac{1}{2k} \big( \phi_p^\beta{'} \sigma_q^* - \sigma_p' \phi_q^{\alpha*} \big) +\frac{1}{2 k^2} \phi_p^\beta \sigma_q^* \nonumber \\
    &=  \frac{1}{2} \phi_p^\beta \phi_q^{\alpha*} + \frac{1}{2k^2} \nabla \phi_q^{\alpha*} \cdot \nabla \phi_p^\beta\,. \label{eqap:4}
\end{align}
Integrating Eq.~\eqref{eqap:4} over $\Omega$ and applying the divergence theorem yields
\begin{align}
    \frac{1}{2k} &\int_{d\Omega_f} \unit{r} \cdot \Big[ \phi_{p}^\beta{'} \nabla \phi_{q}^{\alpha*}  - \phi_{q}^* \nabla \phi_{p}^\beta{'} + \frac{1}{k} \phi_{p}^\beta \nabla \phi_{q}^{\alpha*} \Big] dS \nonumber \\
    &\quad - \frac{1}{2k} \int_{d\Omega_s} \sigma_p' \phi_q^{\alpha*}  dS + \frac{1}{2k^2} \int_{d\Omega_s} \big( k \phi_p^\beta \big)' \sigma_q^* dS \nonumber \\
    &=  \frac{1}{2} \int_{\Omega} \phi_p^\beta \phi_q^{\alpha*} dV 
    + \frac{1}{2k^2} \int_{\Omega} \nabla \phi_p^\beta \cdot \nabla \phi_q^{\alpha*} dV\,. 
    \label{eqap:5}
\end{align}
Therefore
\begin{align}
    \matr{Q}_{qp}^{\alpha,\beta} &= \frac{1}{2} \int_{\Omega} ( \phi_p^\beta \phi_q^{\alpha*} - \phi_{p,\infty}^\beta \phi_{q,\infty}^{\alpha*} ) dV \nonumber \\
    &\quad + \frac{1}{2k^2} \int_{\Omega} ( \nabla \phi_p^\beta \cdot \nabla \phi_q^{\alpha*} - \nabla \phi_{p,\infty}^\beta \cdot \nabla \phi_{q,\infty}^{\alpha*} ) dV \nonumber \\
    &= T_{1,qp}^{\alpha,\beta} + T_{2,qp}^{\alpha,\beta} + T_{S,qp}^{\alpha,\beta} - T_{\infty,qp}^{\alpha,\beta}
    \label{eqap:6}
\end{align}
where
\begin{align}
    T_{1,qp}^{\alpha,\beta} &= -\frac{1}{2k} \int_{d\Omega_s} \phi_q^{\alpha*}\sigma_p' dS\\
    T_{2,qp}^{\alpha,\beta} &= \frac{1}{2k^2} \int_{d\Omega_s} \Big( k \phi_p^{\beta} \Big)'\sigma_q^* dS\\
    T_{S,qp}^{\alpha,\beta} &= \frac{1}{2k} \int_{d\Omega_f} \hat{\vr} \cdot 
    ( \phi_{p,\infty}^{\beta}{'} \nabla \phi_{q,\infty}^{\alpha*} \nonumber \\
    & \quad - \phi_{q,\infty}^{\alpha*} \nabla \phi_{p,\infty}^{\beta}{'}
    + \frac{1}{k}\phi_{p,\infty}^{\beta} \nabla_{q,\infty}^{\alpha*} ) dS\\
    T_{\infty,qp}^{\alpha,\beta} &= \frac{1}{2} \int_{\Omega} \phi_{q,\infty}^{\alpha*} \phi_{p,\infty}^\beta dV \nonumber \\
    & \quad + \frac{1}{2k^2} \int_{\Omega} \nabla \phi_{q,\infty}^{\alpha*} \cdot \nabla \phi_{p,\infty}^\beta dV \,.
\end{align}


\subsection{Computation of $\matr{Q}_{qp}^{inc,inc}$}

$T_{1,qp}^{inc,inc}$ and $T_{2,qp}^{inc,inc}$ vanish since the incident fields $\phi_p^{inc},\phi_q^{inc}$ are source-free, i.e., $\sigma_p=\sigma_q=0$. Therefore,
\begin{align}
    \matr{Q}_{qp}^{inc,inc} &= T_{S,qp}^{inc,inc} - T_{\infty,qp}^{inc,inc} \nonumber \\
    &= \frac{1}{2k} ( I_{1,1} - I_{1,2} + \frac{1}{k}I_{1,3} ) - \frac{1}{2} I_{1,4} - \frac{1}{2k^2} I_{1,5} \label{eq:Q_qp_incinc}
\end{align}
and
\begin{align}
    I_{1,1} &= \int_{d\Omega_f} \phi_{p,\infty}^{inc}{'} \hat{\vr} \cdot 
    \nabla \phi_{q,\infty}^{inc*} dS \nonumber \\
    &= kR \delta_{qp} \Big[2 - (-1)^{l} e^{-2jkR}(-kR+j) + (-1)^{l'} e^{2jkR}(j+kR) \Big] \nonumber \\
    I_{1,2} &= \int_{d\Omega_f} \phi_{q,\infty}^{inc*} \hat{\vr} \cdot \nabla \phi_{p,\infty}^{inc}{'} dS \nonumber \\
    &= -kR \delta_{qp} \Big[2 + (-1)^{l+1} e^{-2jkR} + (-1)^{l'+1} e^{2jkR} \Big] \nonumber \\
    I_{1,3} &= \int_{d\Omega_f} \phi_{p,\infty}^{inc} \hat{\vr} \cdot \nabla \phi_{q,\infty}^{inc*} dS \nonumber \\
    &= \delta_{qp}(-\frac{2}{R}) \nonumber \\
    I_{1,4} &= \int_{\Omega} \phi_{q,\infty}^{inc*} \phi_{p,\infty}^{inc} dV \nonumber \\
    &= \delta_{qp} \Big[ 2R + (-1)^{(l+1)}\frac{\sin{(2kR)}}{k} \Big] \nonumber \\
    I_{1,5} &= \int_{\Omega} \nabla \phi_{q,\infty}^{inc*} \cdot \nabla \phi_{p,\infty}^{inc} dV \nonumber \\
    &= \delta_{qp} 2 k^2 R \nonumber
\end{align}
where the expressions \eqref{eq:phi_inc_approx_start} -- \eqref{eq:phi_inc_approx_end} are made use of and the large argument approximation of $R \rightarrow \infty$ is taken.

Substituting the above results into \eqref{eq:Q_qp_incinc} yields
\begin{align}
    \matr{Q}_{qp}^{inc,inc} &= 0 \,. \nonumber
\end{align}

\subsection{Computation of $\matr{Q}_{qp}^{sca,inc}$}

$T_{1,qp}^{sca,inc}$ vanish since the incident field $\phi_p^{inc}$ is source-free. Therefore, 
\begin{align}
    \matr{Q}_{qp}^{sca,inc} &= T_{2,qp}^{sca,inc} + T_{S,qp}^{sca,inc} - T_{\infty,qp}^{sca,inc} \nonumber \\
    &= T_{2,qp}^{sca,inc} + \frac{1}{2k} ( I_{2,1} - I_{2,2} + \frac{1}{k}I_{2,3} ) \nonumber \\
    & \quad - \frac{1}{2} I_{2,4} - \frac{1}{2k^2} I_{2,5} \,.
    \label{eq:Q_qp_scainc}
\end{align}
Note that
\begin{align}
\int_{d\Omega_f} \sum_t \matr{P}_{tq}^* Y_{lm} Y_{l''m''} dS = (-1)^m \matr{P}_{\hat{p}q}^*
\label{eq:Q_qp_scainc_indentity}
\end{align}
where $p = (l,m)$, $\hat{p} = (l,-m)$, $q = (l',m')$, $t=(l'',m'')$. 
Substitute the expressions \eqref{eq:phi_sca_approx_start} -- \eqref{eq:phi_sca_approx_end} into Eq.~\eqref{eq:Q_qp_scainc}, make use of Eq.~\eqref{eq:Q_qp_scainc_indentity} and take the large argument approximation of $R \rightarrow \infty$, 
\begin{align}
    I_{2,1} &= \int_{d\Omega_f} \phi_{p,\infty}^{inc}{'} \hat{\vr} \cdot 
    \nabla \phi_{q,\infty}^{sca*} dS \nonumber \\
    &= (-1)^{m}\matr{P}_{\hat{p}q}^* (-kR - j) \Big[ e^{2jkR}+(-1)^l \Big] \nonumber \\
    I_{2,2} &= \int_{d\Omega_f} \phi_{q,\infty}^{sca*} \hat{\vr} \cdot \nabla \phi_{p,\infty}^{inc}{'} dS \nonumber \\
    & = (-1)^{m}\matr{P}_{\hat{p}q}^*kR \Big[-e^{2jkR}+(-1)^l\Big] \nonumber \\
    I_{2,3} &= \int_{d\Omega_f} \phi_{p,\infty}^{inc} \hat{\vr} \cdot \nabla \phi_{q,\infty}^{sca*} dS \nonumber \\
    &= (-1)^{m}\matr{P}_{\hat{p}q}^* (jk-\frac{1}{R}) \Big[e^{2jkR}-(-1)^l\Big] \nonumber \\
    I_{2,4} &= \int_{\Omega} \phi_{q,\infty}^{sca*} \phi_{p,\infty}^{inc} dV \nonumber \\
    &= (-1)^{m}\matr{P}_{\hat{p}q}^* \Big[ -R(-1)^l \Big] \nonumber \\
    I_{2,5} &= \int_{\Omega} \nabla \phi_{q,\infty}^{sca*} \cdot \nabla \phi_{p,\infty}^{inc} dV \nonumber \\
    &= (-1)^{m+l+1} \matr{P}_{\hat{p}q}^* k^2 R \,. \nonumber
\end{align}

Substituting the above results into \eqref{eq:Q_qp_scainc} yields
\begin{align}
    \matr{Q}_{qp}^{sca,inc} 
    &= \frac{1}{2k} \int_{d\Omega_s} \phi_p^{inc}{'} \sigma_q^* dS \,. \nonumber
\end{align}

\subsection{Computation of $\matr{Q}_{qp}^{inc,sca}$}

It can been observed from Eq.~\eqref{eqap:6} that
\begin{align}
    \matr{Q}_{qp}^{inc,sca} = \left[ \matr{Q}_{pq}^{sca,inc} \right]^* \,.
\end{align}

\subsection{Computation of $\matr{Q}_{qp}^{sca,sca}$}

\begin{align}
    \matr{Q}_{qp}^{sca,sca} &= T_{1,qp}^{sca,sca} + T_{2,qp}^{sca,sca} + T_{S,qp}^{sca,sca} - T_{\infty,qp}^{sca,sca}
    \label{eq:Q_qp_scasca}
\end{align}
\begin{align}
    T_{1,qp}^{sca,sca} &= -\frac{1}{2k}\int_{d\Omega_s} \phi_q^{sca*} \sigma_p' dS \nonumber \\
    &= -\frac{1}{2k} \int_{d\Omega_s} \int_{d\Omega_s} \sigma_q^*(\vrp') \sigma_p'(\vrp) G^*(\vrp,\vrp') dS'' dS' \nonumber
\end{align}
\begin{align}
    T_{2,qp}^{sca,sca} &= \frac{1}{2k^2}\int_{d\Omega_s} (k\phi_p^{sca})' \sigma_q^* dS \nonumber \\
    &= \frac{1}{2k} \int_{d\Omega_s} \int_{d\Omega_s} \sigma_q^*(\vrp') \sigma_p'(\vrp) G(\vrp,\vrp') dS'' dS' \nonumber \\
    &\quad + \frac{1}{2k} \int_{d\Omega_s} \int_{d\Omega_s} \sigma_q^*(\vrp') \sigma_p(\vrp) \frac{1}{4 \pi} (-j \cos{kD} \nonumber \\
    &\quad\quad\quad \quad\quad - \sin{kD}) dS'' dS' \nonumber \\
    &\quad + \frac{1}{2k^2} \int_{d\Omega_s} \int_{d\Omega_s} \sigma_q^*(\vrp') \sigma_p(\vrp) \frac{\cos{kD}-j\sin{kD}}{4 \pi D} \nonumber
\end{align}
where $D = |\vrp' - \vrp|$ is defined. Adding $T_{1,qp}^{sca,sca}$ and $T_{2,qp}^{sca,sca}$ yields
\begin{align}
    &T_{1,qp}^{sca,sca} + T_{2,qp}^{sca,sca} \nonumber \\ 
    &= \frac{-j}{4 \pi} \int_{d\Omega_s} \int_{d\Omega_s} \sigma_q^*(\vrp') \sigma_p'(\vrp) \frac{\sin{kD}}{kD} dS'' dS' \nonumber \\ 
    &+ \frac{1}{8 \pi k} \int_{d\Omega_s} \int_{d\Omega_s} \sigma_q^*(\vrp') \sigma_p(\vrp) (-j \cos{kD}- \sin{kD}) dS'' dS' \nonumber \\ 
    &+ \frac{1}{8 \pi k} \int_{d\Omega_s} \int_{d\Omega_s} \sigma_q^*(\vrp') \sigma_p(\vrp) \frac{\cos{kD} -j\sin{kD}}{kD} dS'' dS'
\end{align}

\begin{align}
    T_{S,qp}^{sca,sca} &= \frac{1}{2k} (I_{3,1}-I_{3,2}+\frac{1}{k}I_{3,3})
\end{align}
and
\begin{align}
I_{3,1} &= \int_{d\Omega_f} \phi_{p,\infty}^{sca}{'} \hat{\vr} \cdot \nabla \phi_{q,\infty}^{sca*} dS \nonumber \\
&= \frac{jk}{4 \pi} \int_{d\Omega_s} \int_{d\Omega_s} \sigma_q^*(\vrp') \sigma_p'(\vrp) \frac{\sin{kD}}{kD} dS''dS'\nonumber \\
&\quad + \frac{kR}{4 \pi} \int_{d\Omega_s} \int_{d\Omega_s} \sigma_q^*(\vrp') \sigma_p(\vrp) \frac{\sin{kD}}{kD} dS''dS'\nonumber \\
&\quad + \frac{jk}{4 \pi} \int_{d\Omega_s} \int_{d\Omega_s} \sigma_q^{*}(\vrp') \sigma_p(\vrp) \frac{\vrp \cdot \hat{d}}{k^2 D^2} \nonumber \\
&\quad\quad\quad\quad\quad (kD \cos{kD} - \sin{kD}) dS''dS'\nonumber \\
I_{3,2} &= \int_{d\Omega_f} \phi_{q,\infty}^{sca*} \hat{\vr} \cdot \nabla \phi_{p,\infty}^{sca}{'} dS \nonumber \\
&= -\frac{jk}{4 \pi} \int_{d\Omega_s} \int_{d\Omega_s} \sigma_q^*(\vrp') \sigma_p'(\vrp) \frac{\sin{kD}}{kD} dS''dS'\nonumber \\
&\quad -\frac{kR+j}{4 \pi} \int_{d\Omega_s} \int_{d\Omega_s} \sigma_q^*(\vrp') \sigma_p(\vrp) \frac{\sin{kD}}{kD} dS'dS''\nonumber \\
&\quad - \frac{jk}{4 \pi} \int_{d\Omega_s} \int_{d\Omega_s} \sigma_q^*(\vrp') \sigma_p(\vrp) \frac{\vrp \cdot \hat{d}}{k^2 D^2} \nonumber \\
&\quad\quad\quad\quad\quad (kD \cos{kD} - \sin{kD}) dS''dS'\nonumber \\
I_{3,3} &= \int_{d\Omega_f} \phi_{p,\infty}^{sca} \hat{\vr} \cdot \nabla \phi_{q,\infty}^{sca*} dS \nonumber \\
&= \frac{jk}{4 \pi} \int_{d\Omega_s} \int_{d\Omega_s} \sigma_q^*(\vrp') \sigma_p(\vrp) \frac{\sin{kD}}{kD} dS''dS' \nonumber
\end{align}
where $\hat{d} = \frac{\vrp'-\vrp}{|\vrp'-\vrp|}$ is defined. Therefore,
\begin{align}
    T_{S,qp}^{sca,sca} &= \frac{j}{4 \pi} \int_{d\Omega_s} \int_{d\Omega_s} \sigma_q^*(\vrp') \sigma_p'(\vrp) \frac{\sin{kD}}{kD} dS''dS' \nonumber \\
    &\quad+ \frac{R}{4 \pi} \int_{d\Omega_s} \int_{d\Omega_s} \sigma_q^*(\vrp') \sigma_p(\vrp) \frac{\sin{kD}}{kD} dS''dS' \nonumber \\
    &\quad+ \frac{j}{4 \pi k} \int_{d\Omega_s} \int_{d\Omega_s} \sigma_q^*(\vrp') \sigma_p(\vrp) \frac{\sin{kD}}{kD} dS''dS' \nonumber \\
    &\quad+ \frac{j}{4 \pi} \int_{d\Omega_s} \int_{d\Omega_s} \sigma_q^*(\vrp') \sigma_p(\vrp) \frac{\vrp \cdot \hat{d}}{k^2 D^2} \nonumber \\
&\quad\quad\quad\quad\quad (kD \cos{kD} - \sin{kD}) dS''dS'\nonumber
\end{align}

\begin{align}
    T_{\infty,qp}^{sca,sca} &= \frac{1}{2}\int_{\Omega} \phi_{q,\infty}^{sca*} \phi_{p,\infty}^{sca} dV \nonumber \\
    &\quad + \frac{1}{2k^2}\int_{\Omega} \nabla \phi_{q,\infty}^{sca*} \cdot \nabla \phi_{p,\infty}^{sca} dV \nonumber \\
    &= \frac{R}{4 \pi} \int_{d\Omega_s} \int_{d\Omega_s} \sigma_q^*(\vrp') \sigma_p(\vrp) \frac{\sin{kD}}{kD} dS''dS' \,.
\end{align}
Substitute the above results into Eq.~\eqref{eq:Q_qp_scasca},
\begin{align}
    \matr{Q}_{qp}^{sca,sca} &= \frac{1}{4 \pi} \int_{d\Omega_s} \int_{d\Omega_s} \sigma_q^*(\vrp') \sigma_p(\vrp) \Bigg[ \frac{-j \cos{kD} - \sin{kD}}{2k} \nonumber \\
    &\quad + \frac{\cos{kD}+j \sin{kD}}{2k^2 D} +j \vrp \cdot \hat{d} \frac{kD \cos{kD} - \sin{kD}}{k^2 D^2} \Bigg] dS''dS' \nonumber \\
    &= \int_{d\Omega_s} \int_{d\Omega_s} \sigma_q^*(\vrp') \sigma_p(\vrp) \Big[ G_i(\vrp',\vrp) \nonumber \\
    &\quad \quad + G_d(\vrp',\vrp) \Big] dS''dS' \nonumber
\end{align}
where
\begin{align}
G_i(\vrp',\vrp) &= \frac{\cos{kD}}{8 \pi k^2 D} - \frac{\sin{kD}}{8 \pi k} \nonumber \\
&= \frac{1}{2 k^2} \mathbb{R}e\Big[ \frac{d}{dk}(kG(\vrp',\vrp)) \Big] \nonumber \\
G_d(\vrp',\vrp) &= \frac{j \hat{d} \cdot (\vrp' + \vrp)}{8 \pi} \frac{kD \cos{kD}-\sin{kD}}{k^2 D^2} \,. \nonumber
\end{align}

\bibliographystyle{plain}
\bibliography{ref}

\end{document}